\documentclass[apj]{emulateapj}
\usepackage[]{natbib}

\newcommand{\etal}{et al.}

\newcommand{\msun}{\ensuremath{{\rm M}_{\odot}}}

\newcommand{\kms}{km~s\ensuremath{^{-1}}}
\newcommand{\mbh}{\ensuremath{M_\mathrm{BH}}}
\newcommand{\msigma}{\ensuremath{\mbh-\sigma_*}}

\newcommand{\Ha}{H$\alpha$}
\newcommand{\Hb}{H$\beta$}
\newcommand{\oiii}{[O {\sc iii}]}
\newcommand{\oii}{[O {\sc ii}]}
\newcommand{\oi}{[O {\sc i}]}

\shorttitle{Accretion Properties of YRGs}
\shortauthors{SON et al.}

\begin{document}

\title{Accretion Properties of High- and Low-Excitation Young Radio Galaxies}

\author{Donghoon Son\altaffilmark{1}}
\author{Jong-Hak Woo\altaffilmark{1,8}}
\author{Sang Chul Kim\altaffilmark{2}}
\author{Hai Fu\altaffilmark{3}}
\author{Nozomu Kawakatu\altaffilmark{4}}
\author{Vardha N. Bennert\altaffilmark{5}}
\author{Tohru Nagao\altaffilmark{6,7}}
\author{Daeseong Park\altaffilmark{1}}

\affil{$^1$Astronomy Program, Department of Physics and Astronomy, Seoul National University, Seoul 151-742, Republic of Korea}
\affil{$^2$Korea Astronomy and Space Science Institute, Daejeon 305-348, Republic of Korea}
\affil{$^3$Department of Physics and Astronomy, University of California, Irvine, CA 92697, USA}
\affil{$^4$Graduate School of Pure and Applied Sciences, University of Tsukuba, 1-1-1 Tennodai, Tsukuba 305-8571, Japan}
\affil{$^5$Physics Department, California Polytechnic State University, San Luis Obispo, CA 93407, USA}
\affil{$^6$The Hakubi Center for Advanced Research, Kyoto University,
Yoshida-Ushinomiya-cho, Sakyo-ku, Kyoto 606-8302, Japan}
\affil{$^7$Department of Astronomy, Kyoto University, Kitashirakawa-Oiwake-cho, Sakyo-ku, Kyoto 606-8502, Japan}
\altaffiltext{8}{corresponding author; {\tt woo@astro.snu.ac.kr}}

\begin{abstract}

Young radio galaxies (YRGs) provide an ideal laboratory to explore the connection 
between accretion disk and radio jet thanks to their recent jet formation. 
We investigate the relationship between the emission-line properties, the black hole 
accretion rate, and the radio properties using a sample of 34 low-redshift ($z < 0.4$) YRGs. 
We classify YRGs as high-excitation galaxies (HEGs) and low-excitation galaxies (LEGs) 
based on the flux ratio of high-ionization to low-ionization emission lines. 
Using the \Ha\ luminosities as a proxy of accretion rate, 
we find that HEGs in YRGs have $\sim1$ dex higher Eddington ratios than LEGs in YRGs, 
suggesting that HEGs have higher mass accretion rate or higher radiative efficiency than LEGs. 
In agreement with previous studies, we find that the luminosities of emission lines, in 
particular \Ha, are correlated with radio core luminosity,
suggesting that accretion and young radio activities are fundamentally connected. 

\end{abstract}


\section{INTRODUCTION}

Active galactic nuclei (AGN)  play an important role in galaxy evolution
by feeding their host galaxies with radiative and/or kinetic energy,
leading to the observed scaling relations between black hole (BH) mass 
and galaxy properties (e.g., Ferrarese \& Merritt 2000; G{\"u}ltekin \etal~2009; Woo \etal~2010).
Thus, investigating the radiative and kinetic processes of AGN is of fundamental importance 
for understanding AGN physics as well as feedback mechanisms.

The formation of relativistic jets and its connection to accretion disk 
remain as an open issue in AGN physics 
(e.g., Rees 1984; Meier 2003; McKinney 2006; Komissarov \etal~2007; McKinney \etal~2012).
Thanks to their short dynamical timescale, X-ray binaries in various states have 
been observed in detail, revealing that individual sources occupy a particular 
accretion state with various X-ray and radio luminosities (e.g., Fender \etal~2004; 
Remillard \& McClintock 2006).
By analogy to X-ray binaries, the power of AGN jets may also depend on the physical states of accretion disk.

However, the disk-jet connection is more complicated in AGNs.
It has been known that the radio power is correlated with narrow emission 
line luminosities, i.e., \oiii, which is a proxy for the accretion power 
(e.g., Baum \& Heckman 1989; Rawlings \& Saunders 1991), 
indicating that the jet launching mechanism is connected with accretion disk. 
However, the correlation shows a large scatter (e.g., Morganti \etal~1992; Labiano 2008),
implying that the nature of the disk-jet connection is complicated and that
other physical parameters are necessary to constrain.

The properties of the accretion disk seem to play an additional role in the disk-jet connection.
At a given radio luminosity, high-excitation galaxies (HEG) classified with
high \oiii/\Ha\ ratio (e.g., Laing \etal~1994) have an order of magnitude 
higher \oiii\ luminosity than low-excitation galaxies (LEG).
Using a sample of 3C radio galaxies, 
Buttiglione \etal~(2010) showed that there are two sequences in the radio--emission 
line luminosity plane, suggesting that systematically different accretion rates or accretion modes 
are responsible for the separation between HEGs and LEGs.

One of the limiting factors in interpreting the disk-jet connection in 
powerful radio galaxies is that the lifetime of large radio sources
is much longer than the transition timescale of the physical states 
in the accretion disk (O'Dea \etal~2009). Thus, comparing radio and 
accretion powers in radio galaxies with large-scale jets may suffer
systematic uncertainties, leading to large scatters in the correlation
between optical and radio properties (see Punsly \& Zhang 2011).

In contrast, compact radio galaxies with small jets triggered by recent 
activity are very useful to investigate the disk-jet connection,
since radio and disk activities are contemporaneous. 
The ages of compact radio galaxies are typically estimated as $t_{\rm age}
=10^{2}-10^{3}$ years  (e.g., Orienti \etal~2007; Fanti 2009; Giroletti \& Polatidis 2009),   
while the lifetime of extended (up to a few hundred kpc) radio sources are 
about $10^{6}-10^{7}$ years (e.g., Alexander \& Leahy 1987; Carilli \etal~1991; Fanti \etal~1995;
O'Dea \etal~2009).

Recently, a substantial amount of such compact radio galaxies have been detected 
and classified with various characteristics:
compact symmetric objects (CSO) with a linear scale $\lesssim 1$ kpc, 
gigahertz-peaked spectrum (GPS; $\lesssim 1$ kpc) sources,
medium-size symmetric objects ($1-10$ kpc), 
and compact steep-spectrum (CSS;  $\lesssim 20$ kpc) sources (See O'Dea \etal~2009).
These young radio galaxies (YRGs) are a good laboratory to investigate
the physical link between AGN jets and accretion disks.

YRGs show a close connection between emission-line gas and radio properties.
For example, it has been reported that emission-line gas (e.g.,\oii, \oiii) is well 
aligned with the radio jet (e.g., de Vries \etal~1999; Axon \etal~2000) and that 
emission-line luminosities (e.g., \oiii) are correlated with the radio power (e.g., Labiano 2008;  Buttiglione \etal~2010; Kunert-Bajaraszewska \& Labiano 2010).
The optical emission-line diagnostics of the narrow-line region (NLR) can constrain
accretion properties, since the NLR is photoionized by the nuclear continuum radiation (e.g., Kawakatu \etal~2009). 

In this work, using a sample of YRGs covering a large range of luminosities, 
we investigate the properties of NLRs and accretion 
by directly measuring narrow emission-line luminosities from optical spectra, 
and compare them with radio properties for constraining the disk-jet connection. 
In Section~\ref{obs}, we describe the sample selection, spectroscopic observations, 
data reduction, and radio data collection. 
The measurements of emission-line fluxes and the stellar velocity dispersions are
described in Section~\ref{analysis}.
Main results are presented in Section~\ref{result} and Section~\ref{sum} contains discussions and summary. 
Throughout the paper, we used cosmological parameters, $H_0 = 70$ \kms~ Mpc$^{-1}$, $\Omega_{m} = 0.3$, and $\Omega_{\Lambda} = 0.7$.

\begin{deluxetable*}{llllllllrcccccr}                                                                                                                                            
\tablewidth{0pt}
\tabletypesize{\scriptsize}
\tablecaption{Sample properties}
\tablehead{ 
\colhead{Name} & \colhead{R.A.} & \colhead{Decl.} & \colhead{z} & \colhead{$E(B-V)$} & \multicolumn{3}{c}{AGN type}                                                                        
  & \colhead{$S_{1.4}$}  & \colhead{Ref.} & \colhead{Jet size} & \colhead{Ref.}  & \colhead{Run} & \colhead{Exposure} & \colhead{S/N}\\                                          
\colhead{} & \multicolumn{2}{c}{(J2000)} &  \colhead{} &  \colhead{}   &  \colhead{} &  \colhead{}                                                                               
  &  \colhead{} &  \colhead{(Jy)} &  \colhead{}&  \colhead{(kpc)} &  \colhead{} &  \colhead{}  & \colhead{(hr)} & \colhead{} \\                                                  
\colhead{(1)} & \colhead{(2)} &  \colhead{(3)} &  \colhead{(4)} &  \colhead{(5)} &  \colhead{(6)} &  \colhead{(7)}                                                               
  &  \colhead{(8)} &  \colhead{(9)} &  \colhead{(10)} &  \colhead{(11)} & \colhead{(12)} & \colhead{(13)} & \colhead{(14)}  & \colhead{(15)}                                     
}                                                                                                                                                                                
\startdata
\multicolumn{15}{c}{Lick and Palomar targets}\smallskip \\
\hline                                                                                                                                                                      
0019$-$000  	& 00:22:25  & +00:14:56				& 0.305		&	0.027				&	2		  	 &	GPS			&	HEG			& 2.919 	 & F   	& 0.121	 & O98			&	1 		 & 1.4			& 48   \\  
0035+227 			& 00:38:08  & +23:03:28				& 0.096		&	0.033				&	2		  	 &	CSO			&	LEG			& 0.547  	 & N   	& 0.015	 & K08			&	3 		 & 2				& 73   \\  
0134+329 			& 01:37:41  & +33:09:35				& 0.367		&	0.044				&	1	  	   &	CSS	  	&	HEG			& 16.018	 & N   	& 1.100  & O98			&	4 		 & 1				& 208  \\  
0221+276 			& 02:24:12  & +27:50:12				& 0.310		&	0.125				&	1		  	 &	CSS			&	HEG			& 3.024 	 & N   	& 5.536  & K08			&	4 		 & 1.5			& 32   \\  
0316+413 			& 03:19:48  & +41:30:42				& 0.018		&	0.163				&	1	  	   &	CSO	  	&	HEG			& 22.829	 & N   	& 0.005  & K08			&	3 		 & 1				& 169  \\	 
0345+337 			& 03:48:47  & +33:53:15				& 0.243		&	0.389				&	2		  	 &	CSS			&	LEG			& 2.365 	 & N   	& 0.443  & O98			&	5 		 & 3				& 11   \\  
0428+205 			& 04:31:04  & +20:37:34				& 0.219		&	0.542				&	2		  	 &	GPS			&	LEG			& 3.756 	 & N   	& 0.412  & O98			&	2 		 & 1.5			& 16   \\  
0605+480 			& 06:09:33  & +48:04:15				& 0.277		&	0.162				&	2		  	 &	CSS			&	LEG			& 4.133 	 & N   	& 15.17	 & K08			&	5 		 & 3				& 12   \\  
0941$-$080		& 09:43:37  & $-$08:19:31			& 0.228		&	0.029				& 2  			 &	GPS			&	LEG		  & 2.756 	 & N   	& 0.083	 & O98			&	6 		 & 1.5			& 10   \\  
1203+645 			& 12:06:25  & +64:13:37				& 0.372		&	0.017				&	1		  	 &	CSS			&	HEG			& 3.719  	 & N   	& 3.067	 & K08			&	10 		 & 3				& 21   \\  
1225+442 			& 12:27:42  & +44:00:42				& 0.348		&	0.019				&	2		  	 &	GPS			&	HEG			& 0.383 	 & F   	& 0.493	 & K08			&	7 		 & 1.5			& 9    \\  
1233+418 			& 12:35:36  & +41:37:07				& 0.250		&	0.022				&	2		  	 &	CSS			&	LEG			& 0.664 	 & F   	& 6.068	 & K10			&	7 		 & 0.5			& 8    \\  
1245+676 			& 12:47:33  & +67:23:16				& 0.107		&	0.021				&	2		  	 &	CSO 		&	LEG			& 0.263 	 & N   	& 0.007	 & K08			&	7 		 & 1.5			& 141  \\  
1250+568 			& 12:52:26  & +56:34:20				& 0.320		&	0.010				&	1		  	 &	CSS			&	HEG			& 2.442 	 & F   	& 3.593	 & K08			&	8 		 & 1.5			& 67   \\  
1323+321 			& 13:26:17  & +31:54:10				& 0.368		&	0.015				&	2		  	 &	GPS			&	HEG			& 4.747 	 & F   	& 0.149	 & K08			&	10 		 & 1.5			& 34   \\  
1404+286 			& 14:07:00  & +28:27:15				& 0.077		&	0.018				&	1		  	 &  GPS			&	HEG			& 0.830 	 & F   	& 0.005	 & K08			&	9 		 & 1				& 53   \\	 
1807+698 			& 18:06:51  & +69:49:28				& 0.051		&	0.036				&	2  			 &	CSS 		&	LEG			& 1.886 	 & N   	& 1.726	 & G94			&	3 		 & 1.5			& 120  \\  
1943+546 			& 19:44:32  & +54:48:07				& 0.263		&	0.162				&	2		  	 &	CSO			&	\nodata	& 1.754 	 & N   	& 0.072	 & K08			&	2 		 & 1.5			& 13   \\  
2352+495 			& 23:55:09  & +49:50:08				& 0.238		&	0.181				&	2		  	 &	GPS			&	LEG			& 2.306 	 & N   	& 0.092	 & K08			&	2 		 & 2				& 21 \\   
\cutinhead{SDSS targets}
0025+006 			&	00:28:33  & +00:55:11			  & 0.104		& 0.024				&	2		  	 &	CSS			&	HEG			&	0.237 	 & F   	& 2.230	 & K10			&	     	 & \nodata	& 21    \\         
0754+401 			&	07:57:57  & +39:59:36			  & 0.066		& 0.054				&	2		  	 &  CSS			&	HEG			&	0.099 	 & F   	& 0.178	 & K10			&	     	 & \nodata	& 28    \\         
0810+077 			&	08:13:24  & +07:34:06			  & 0.112		& 0.022				&	2  			 &	CSS			&	LEG			&	0.463 	 & F   	& 1.982	 & K10			&	     	 & \nodata	& 19    \\         
0921+143 			&	09:24:05  & +14:10:22			  & 0.136		& 0.029				&	2		  	 &	CSS			&	LEG			&	0.108    & F   	& 0.520	 & K10			&	     	 & \nodata	& 19    \\         
0931+033 			&	09:34:31  & +03:05:45			  & 0.225		& 0.033				&	2		  	 &	CSS			&	LEG			&	0.292 	 & F   	& 1.140	 & K10			&	     	 & \nodata	& 22    \\         
0942+355 			&	09:45:26  & +35:21:03			  & 0.208		& 0.011				&	1  			 &	CSS			&	HEG			&	0.148  	 & F   	& 3.138	 & K10			&	     	 & \nodata	& 20    \\         
1007+142 			&	10:09:56  & +14:01:54 			& 0.213		& 0.043				&	2		  	 &	CSS			&	LEG			&	1.045 	 & F   	& 2.320	 & K10			&	     	 & \nodata	& 18    \\         
1037+302 			&	10:40:30  & +29:57:58			  & 0.091		& 0.019				&	2		  	 &	CSS			&	LEG			&	0.388  	 & F   	& 2.591	 & K10			&	     	 & \nodata	& 30    \\         
1154+435 			&	11:57:28  & +43:18:06			  & 0.230		& 0.013				&	1		  	 &	CSS			&	HEG			&	0.256 	 & F   	& 3.227	 & K10			&	     	 & \nodata	& 21    \\         
1345+125 			&	13:47:33  & +12:17:24			  & 0.122		& 0.034				&	1 			 &	GPS			&	HEG			&	4.860  	 & F   	& 0.085	 & O98			&	     	 & \nodata	& 17    \\         
1407+363 			&	14:09:42  & +36:04:16			  & 0.148		& 0.012				&	2		  	 &	CSS			&	HEG			&	0.143 	 & F   	& 0.050	 & K10			&	     	 & \nodata	& 9     \\         
1521+324 			&	15:23:49  & +32:13:50			  & 0.110		& 0.026				&	2  			 &	CSS			&	HEG			&	0.169    & F   	& 0.285	 & K10			&	     	 & \nodata	& 17    \\         
1558+536 			&	15:59:28  & +53:30:55			  & 0.179		& 0.012				&	2		  	 &	CSS			&	LEG			&	0.182    & F   	& 3.630	 & K10			&	     	 & \nodata	& 17    \\         
1601+528 			&	16:02:46  & +52:43:58			  & 0.106		& 0.019				&	1		  	 &	CSS			&	LEG			&	0.576 	 & F   	& 0.269	 & K10			&	     	 & \nodata	& 26    \\         
1610+407 			&	16:11:49  & +40:40:21 			& 0.151		& 0.008				&	2		  	 &	CSS			&	LEG			&	0.553 	 & F   	& 1.858	 & K10			&	     	 & \nodata	& 12               
\enddata                                                                                                                                                                         
\tablecomments{Columns: 
(1) Target name; 
(2) R.A.;
(3) Decl..;
(4) Redshift; 
(5) Galactic extinction;                                                                                                                         
(6) Spectroscopic AGN type -- 1: Type 1 AGN with broad emission lines; 
2: Type 2 AGN without broad emission line;
(7) YRG type;                                                                             
(8) Classification by excitation -- HEG: high excitation galaxy, LEG: low excitation galaxy; 
(9) 1.4 GHz integrated flux density of radio source;                                                                                                                                             
(10) Reference for the flux density -- F: from the Faint Images of the Radio Sky at Twenty-centimeters (FIRST) catalog,                                                          
N: from the NRAO/VLA Sky Survey (NVSS) catalog;                                                                    
(11) Jet size;
(12) Reference for the jet size -- G94: Gelderman \& Whittle (1994), O98: O'Dea (1998), K08: Kawakatu \etal~(2008);
K10: Kunert-Bajaraszewska \etal~(2010);                                                                                                                                          
(13) Observing run in Table~\ref{setup};  (14) Exposure time; (15) Signal-to-noise ratios near 5100 \AA\ in the rest frames.}                                                    
\label{targets}                                                                                                                                                                  
\end{deluxetable*}

\section{OBSERVATIONS AND DATA}{\label{obs}}

\subsection{Sample Selection}

To investigate the properties of optical emission lines and their connection
to radio activities, we selected a sample of 34 YRGs from the literature.
Initially, we selected 19 known YRGs listed by O'Dea (1998) and 
Kawakatu \etal~(2008), by limiting a redshift range $z < 0.4$.
This choice was made in order to include narrow emission lines from  
[\ion{O}{2}] to [\ion{Ar}{3}] ($3727-7136$ \AA\ in the rest-frame) in the observed
wavelength range for comparing emission-line flux ratios with the photoionization models. 

During our observing campaign, new YRGs with relatively low luminosity 
have been reported by Kunert-Bajaraszewska \etal~(2010) 
based on the unresolved and isolated morphology in Faint Images of the Radio Sky at Twenty-centimeters (FIRST).
To enlarge the sample size and the luminosity range, we selected 
additional 15 YRGs at $z < 0.3$ from their list, for which
optical spectra with the same rest-frame wavelength range ($3727-7136$ \AA)
were available in the archive of the Sloan Digital Sky Survey (SDSS) Data Release 
7 (DR7; Abazajian \etal~2009). 
Thus, combining new data with SDSS archival data, we compiled a sample of 34 
low-redshift YRGs for investigating optical and radio properties.
Table~\ref{targets} lists the properties of individual YRGs, including optical and radio
AGN classifications.

\subsection{New Optical Data}

Among 34 objects in the sample, SDSS spectra are available for 15 objects,
while for the remaining 19 objects, new spectroscopic data have been obtained 
from the Lick and Palomar telescopes. In this section, we describe new observations and data reduction.

\subsubsection{Observations}

We observed 19 YRGs to obtain high-quality optical spectra. 
Ten objects were observed with the Kast double spectrograph 
at the Lick 3 m telescope (Miller \& Stone 1993). 
We used the D55 dichroic beam splitter to pass the light into blue and red side 
detecters. Depending on the redshift of the target, different gratings were used 
in the red spectrograph, covering the wavelength range of $5500-10000$ \AA~ 
with a wavelength scale of $1.7-4.6$ \AA~pixel$^{-1}$  and spatial scale of $0.78$  
arcsec pixel$^{-1}$. 
For the blue spectrograph, the 600/4310 grism was used, 
covering the wavelength range $3600-5500$ \AA~with a wavelength scale of 
$1.0$ \AA~pixel$^{-1}$ and spatial scale of $0.43$ arcsec pixel$^{-1}$.

The other nine galaxies were observed withthe Double Spectrograph for the Palomar 200-inch Telescope  
(DBSP; Oke \& Gunn 1982) using the D55 or D68 
dichroic mirrors depending on the redshift of the targets. 
The 158/7500 and 316/7500 gratings were used on the red spectrograph, 
covering  $5200-10000$ \AA\ with a wavelength scale of 4.8 or 2.4 \AA\ pixel$^{-1}$ 
and spatial scale of $0.62$ arcsec pixel$^{-1}$, while the 600/4000 grating was used 
on the blue spectrograph, covering the wavelength range of $3200-5500$ \AA\ 
or $4000-6800$ \AA\ with a wavelength scale of $\sim 1.1$ \AA\ pixel$^{-1}$ and spatial 
scale of $0.39$ arcsec pixel$^{-1}$.

For all observations, a 2\arcsec-wide slit was centered on 
the nucleus of the targets after aligned with a parallactic angle.
Several flux photometric standard stars, namely, BD+262606, BD+284211,
Feige 34, G138-31, G191B2B, HD192281, or Wolf1346 were observed during each night. 
A0V stars with similar airmass and hour angle to each target were observed
for telluric correction. 
The properties of each YRG along with exposure time and S/N are presented in Table~\ref{targets}, 
while the details of instrumental setup, observing date, and sky conditions are 
listed in Table~\ref{setup} .

\begin{figure*}
\center
\epsscale{1.1}\plottwo{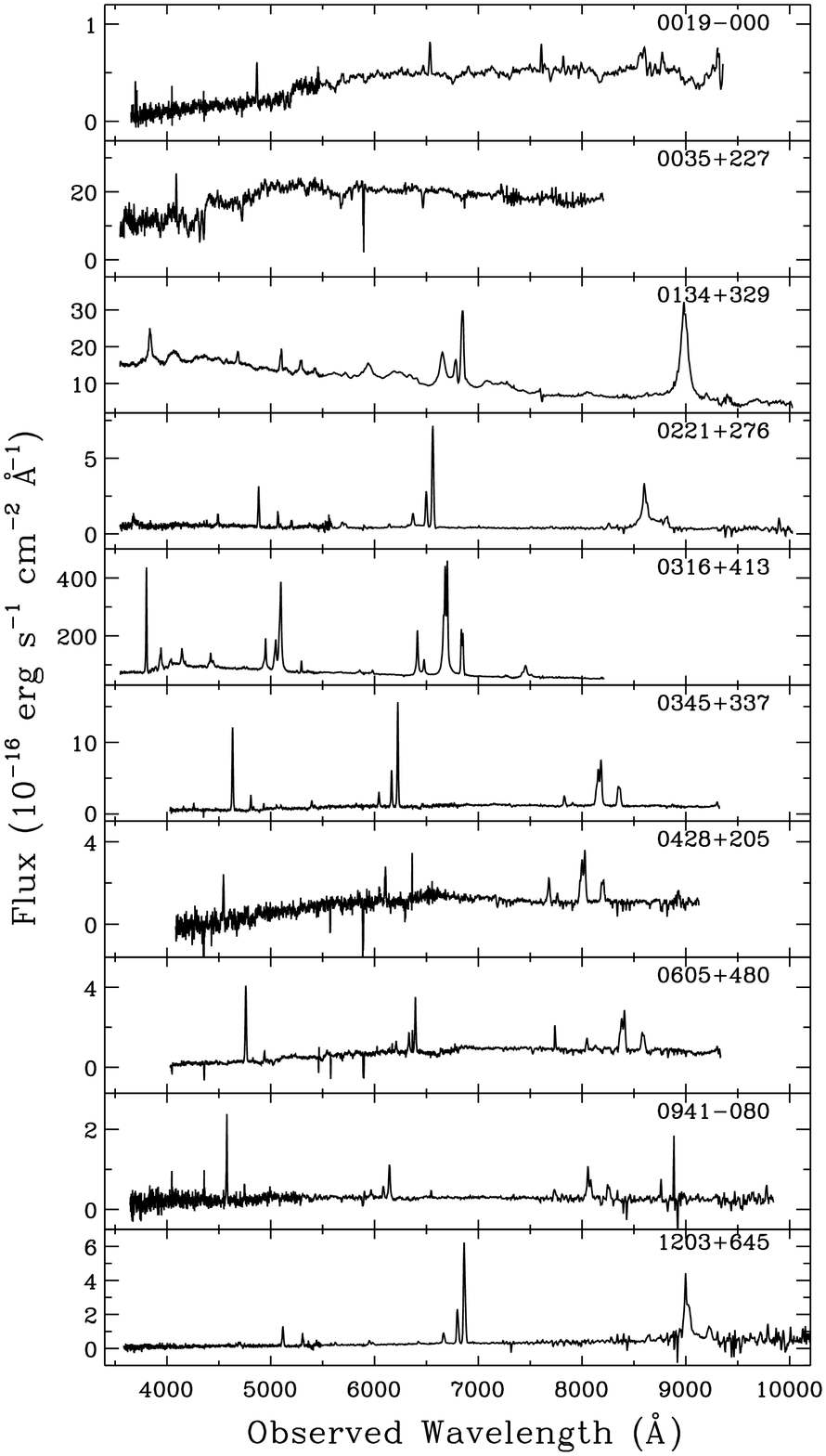}{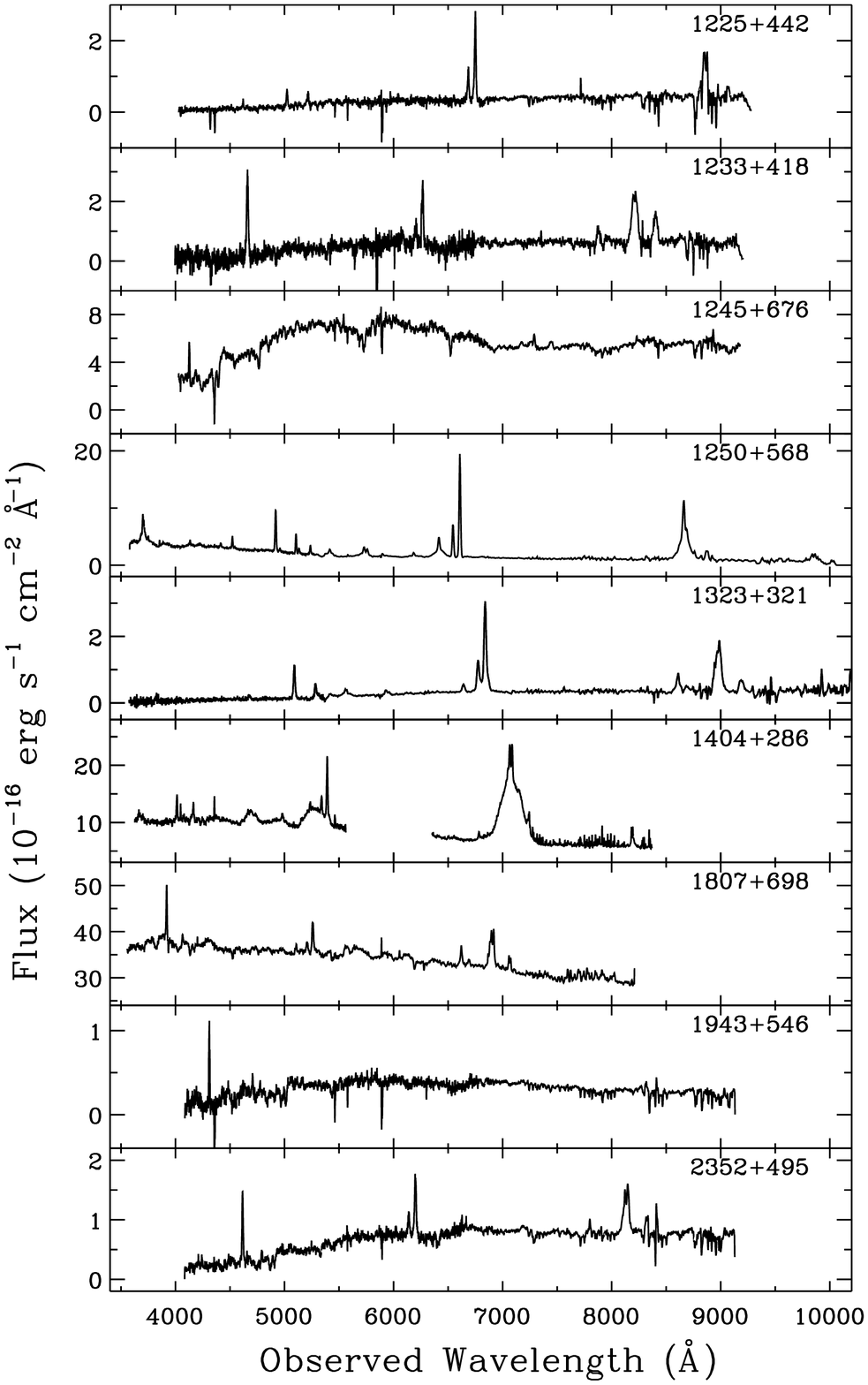}
\caption{Reduced spectra of 19 YRGs obtained with the Kast at the Lick telescope or the DBSP
at the Palomar telescope.
}
\label{spec1}
\end{figure*}

\begin{figure}
\center
\epsscale{1.2}\plotone{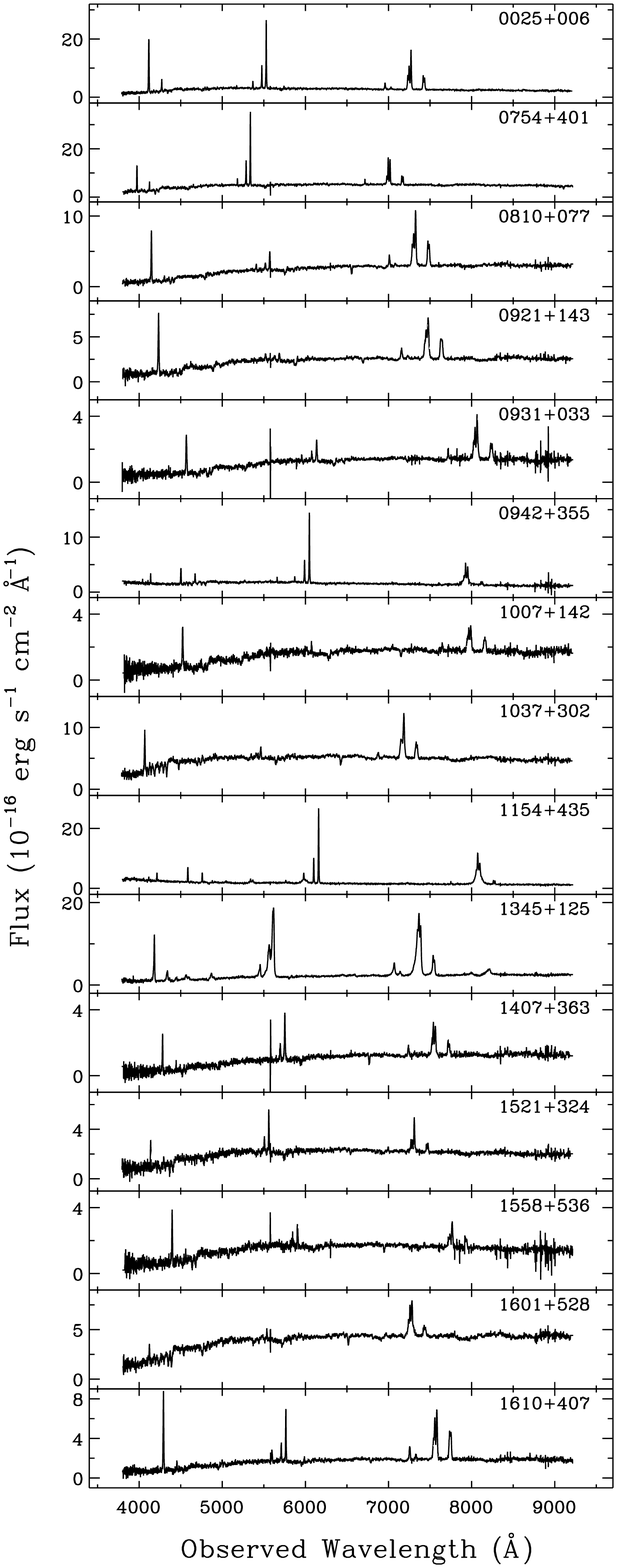}
\caption{Spectra of 15 additional YRGs obtained from the SDSS DR7.
}
\label{spec2}
\end{figure}

\subsubsection{Data Reduction}

Standard spectroscopic data reduction procedures, including bias subtraction, 
flat-fielding, spectral extraction, wavelength calibration, and flux calibration,
were carried out for each data set using a series of IRAF scripts developed for 
long-slit data (e.g., Woo \etal~2005, 2006). 
Telluric A and B bands were corrected 
using an A0V star spectrum obtained for each target as similarly performed
in the previous studies (e.g., Woo \etal~2006). 
After combining multiple exposures in each blue and red sides, the blue- 
and red-side spectra were merged into one final spectrum for each target. 
Galactic extinction corrections were applied using the extinction law
by Cardelli \etal~(1989) with the $E(\bv)$ values listed in Table~\ref{targets}.
We measured the spectral resolution by averaging the width of 
sky or arc lines for each instrumental setup (see Table~\ref{setup}).
In the case of the SDSS archival data, the spectra were obtained with a 3\arcsec\ 
diameter fiber and a spectral resolution of $\lambda/\Delta\lambda \approx 1800$.

\subsection{Radio Data}

To compare with optical spectroscopic properties, 
we collected the radio luminosity and the jet size of the sample 
from the literature, after correcting for the adopted cosmological parameters.
The 1.4 GHz integrated flux densities of radio core were extracted from the FIRST catalog,
 while the NRAO/VLA Sky Survey (NVSS) catalog was used if FIRST data were not available. 
We confirmed that these integrated flux densities were measured from radio core 
by checking the presence of unresolved radio cores in the FIRST and NVSS images. 
The integrated flux densities and their references are listed in Table~\ref{targets}. 

The jet size is generally defined by a half of the linear scale or the hot spot 
distance from the core. 
We collected  the linear jet sizes from the literature after correcting for the 
adopted cosmological parameters. The radio luminosity and the jet size of each object are
listed in Table~\ref{targets}.

\section{ANALYSIS}{\label{analysis}}

In this section, we present AGN emission and stellar absorption line
measurements based on the optical spectra. First, we describe AGN
classification based on the presence of the broad \Ha\ line in Section~\ref{spectra}. 
Second, we present the emission-line fitting procedure and measurements 
in Section~\ref{measure}. Then, we describe 
the stellar velocity dispersion measurements in Section~\ref{svd-sec}.

\subsection{The Spectra}{\label{spectra}}

\begin{figure}
\center
\includegraphics[width=.4\textwidth, height=.8\textwidth]{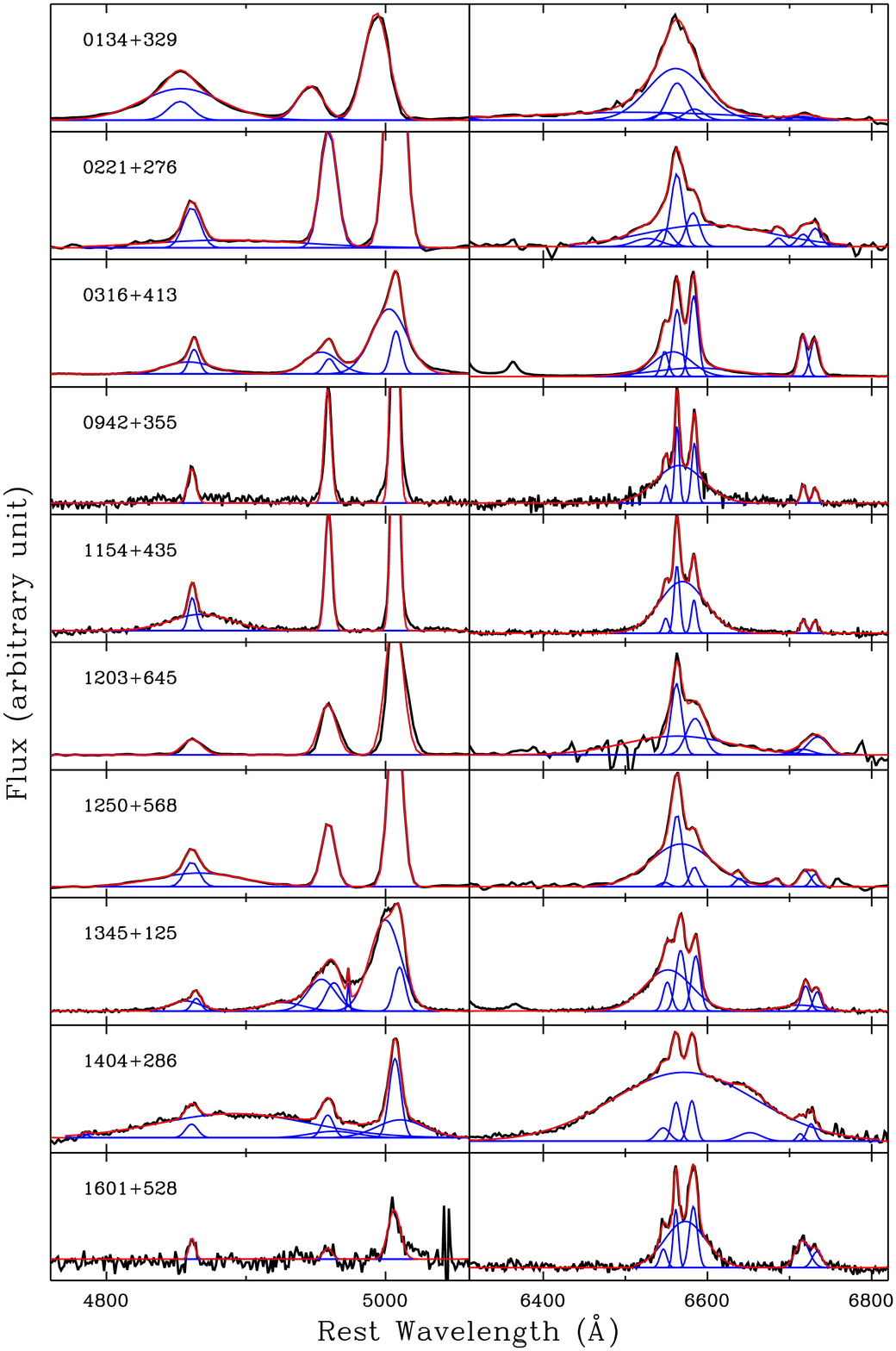}
\caption{\Hb\ and \Ha\ regions of Type 1 AGNs. Broad and narrow components of \Hb\ and \Ha\
lines and narrow forbidden lines, i.e., \oiii, [N {\sc ii}] and [S {\sc ii}] in the continuum-subtracted
spectra (black lines) were simultaneously fitted with Gaussian models (blue). The combined fit
is represented by red lines.
(A color version of this figure is available in the online journal.)}
\label{specfit2}
\end{figure}

In Figure~\ref{spec1},  we present the spectra of 19 YRGs obtained from the Lick and 
Palomar telescopes. For completeness, we also present SDSS spectra of 15 
additional YRGs in Figure~\ref{spec2}.
We classified YRG as Type 1 and Type 2 based on the presence of the broad \Ha\ line.
The sample is composed of 10 Type 1 (i.e., broad-line) and 24 Type 2 (i.e., narrow-line)
AGNs as listed in Table~\ref{targets}. 
For Type 1 objects, the broad H$\alpha$ component
was clearly present although the broad \Hb\ were not detected for three objects,
presumably due to the weak flux level of the broad \Hb\ component 
(for details, see Figure~\ref{specfit2}).
In Type 2 objects, stellar absorption lines were dominant as expected, confirming that
subtracting stellar absorption lines is important to precisely measure AGN emission
line fluxes.
While one object {\it 0810+077} was previously classified as Type 1 AGN,
the broad \Ha\ was not detected in our spectra, hence we classified it as a Type 2 AGN.

The optical spectra of {\it 0428+205, 1203+645, 1225+442, 1233+418}, and {\it 1323+321} 
are newly presented in this paper. The spectra of other objects have higher spectral 
resolution and/or better S/N than spectra presented by previous studies 
(e.g., Gelderman \& Whittle 1994: {\it 0134+329, 0221+276, 0345+337}, and {\it 1250+568};  
Labiano \etal~2005 and O'Dea \etal~2002: {\it 0221+276} and {\it 1250+568}; 
Buttiglione \etal~2009: {\it 0345+337, 0605+480}, and {\it 1807+698}). 

\setlength{\tabcolsep}{0.5pt}

\begin{deluxetable*}{clccccccccccrcl}
\tablecolumns{15}
\tablewidth{0pt}
\tabletypesize{\scriptsize}
\tablecaption{Journal of observations}
\tablehead{
\colhead{Run} & \colhead{Date} & \colhead{Tel.} & \colhead{Inst.} &  \colhead{Dichr.}
& \multicolumn{4}{c}{Blue side}   & \multicolumn{4}{c}{Red side}  
&  \colhead{Seeing} &  \colhead{Sky} \\  
\colhead{} & \colhead{} & \colhead{} & \colhead{} & \colhead{}                                                                        
	&  \colhead{Grism} &  \colhead{Plate} &  \colhead{Spatial} &  \colhead{Res.}                                                                              
	&  \colhead{Grating} &  \colhead{Plate} &  \colhead{Spatial} &  \colhead{Res.}                                                                            
         & \colhead{} & \colhead{} \\                                                                                                        
\colhead{} & \colhead{} & \colhead{} & \colhead{} & \colhead{}                                                                                               
	& \colhead{(l mm$^{-1}$)} & \colhead{(\AA~pixel$^{-1}$)} & \colhead{(arcsec pixel$^{-1}$)} & \colhead{(\AA)}                                                                      
	& \colhead{(l mm$^{-1}$)} & \colhead{(\AA~pixel$^{-1}$)} & \colhead{(arcsec pixel$^{-1}$)} & \colhead{(\AA)}                                                                      
	& \colhead{(arcsec)} &  \colhead{}  \\                                                                                                                      
\colhead{(1)} & \colhead{(2)} &  \colhead{(3)} &  \colhead{(4)}                                                                                               
	&  \colhead{(5)} &  \colhead{(6)} &  \colhead{(7)} &  \colhead{(8)}                                                                                         
	&  \colhead{(9)} &  \colhead{(10)} &  \colhead{(11)} & \colhead{(12)}                                                                                       
	&  \colhead{(13)} & \colhead{(14)}   & \colhead{(15)}                                                                                                       
}                                                                                                                                                             
\startdata                                                                                                                                                    
1  & 2009 Jul 21 & P & D & D55& 600 & 1.1 & 0.39 & 5.0 & 158 & 4.8 & 0.62 & 16.3 & $\sim$2 & clear \\ 
2  & 2009 Aug 23 & P & D & D68& 600 & 1.1 & 0.39 & 5.0 & 316 & 2.4 & 0.62 &  8.1 & $\sim$2.2 & clear \\ 
3  & 2009 Aug 23 & L & K & D55& 600 & 1.0 & 0.43 & 4.4 & 600 & 2.4 & 0.78 &  6.1 & $\sim$1.3 & clear \\ 
4  & 2009 Aug 24 & L & K & D55& 600 & 1.0 & 0.43 & 4.4 & 300 & 4.6 & 0.78 & 13.0 & $\sim$1.5 & clear \\ 
5  & 2009 Dec 09 & P & D & D68& 600 & 1.1 & 0.39 & 5.0 & 316 & 2.4 & 0.62 &  8.1 & $\sim$2 & thick \\ 
6  & 2009 Dec 16 & L & K & D55& 600 & 1.0 & 0.43 & 4.4 & 300 & 4.6 & 0.78 & 13.0 & $\sim$1.5 & thin  \\ 
7  & 2010 Jan 14 & P & D & D68& 600 & 1.1 & 0.39 & 5.0 & 316 & 2.4 & 0.62 &  8.1 & $\sim$3 & thin \\  
8  & 2010 Mar 10 & L & K & D55& 600 & 1.0 & 0.43 & 4.4 & 300 & 4.6 & 0.78 & 13.0 & $\sim$1.5 & humid \\ 
9  & 2011 Jan 06 & L & K & D55& 600 & 1.0 & 0.43 & 4.4 & 830 & 1.7 & 0.78 &  4.2 & $\sim$1.5 & clear \\ 
10 & 2011 Jan 07 & L & K & D55& 600 & 1.0 & 0.43 & 4.4 & 300 & 4.6 & 0.78 & 13.0 & $\sim$2 & clear            
\enddata                                                                                                     
\tablecomments{Columns: (1) Observing run; (2) Observation date (UT); (3) Telescope (P: Palomar 5 m, L: Lick 3 m);
(4) Spectrograph (D: DBSP, K: Kast); (5) Dichroic mirrors;                                                     
(6) \& (10) Grism for the blue side and Grating for the red side, respectively; (7) \& (11) Plate scales for the blue and red sides, respectively;                          
(8) \& (12) Spatial scales for the blue and red sides, respectively;                                                                                          
(9) \& (13) Instrumental resolution; (14) Seeing; (15) Sky condition.}                                                                                        
\label{setup}                                                                                                                                                 
\end{deluxetable*}                                                                                                                                            
\setlength{\tabcolsep}{5pt}

\begin{figure}
\center
\includegraphics[width=0.240\textwidth, height=.8\textwidth]{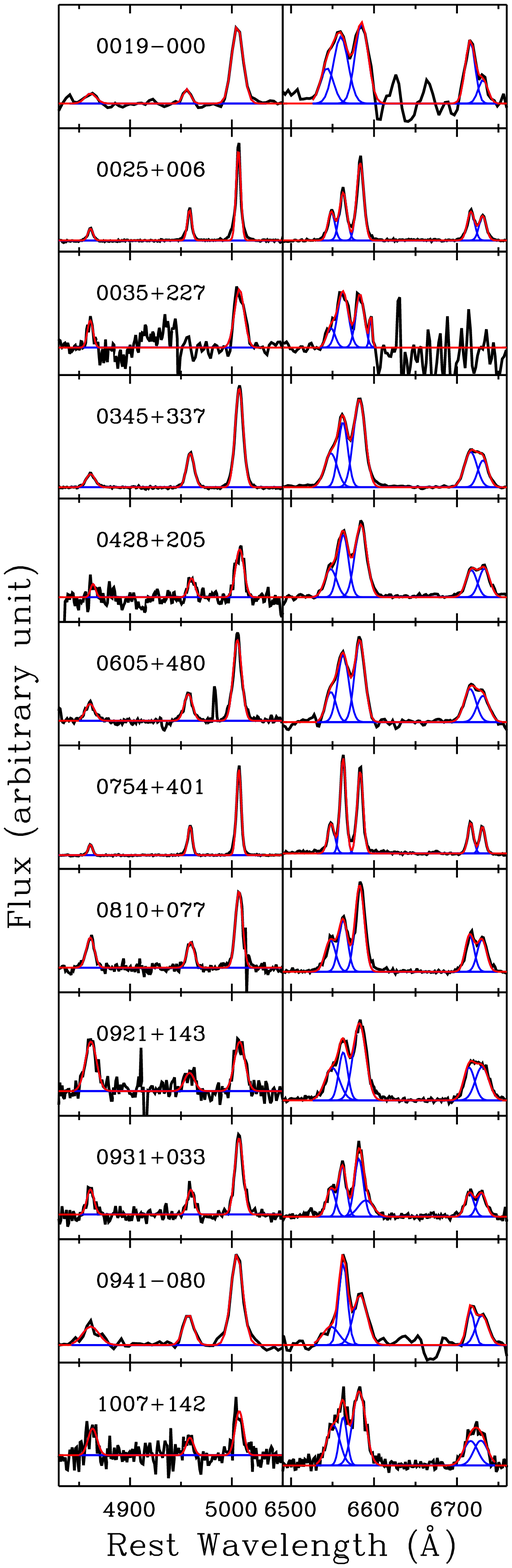}
\includegraphics[width=0.232\textwidth, height=.8\textwidth]{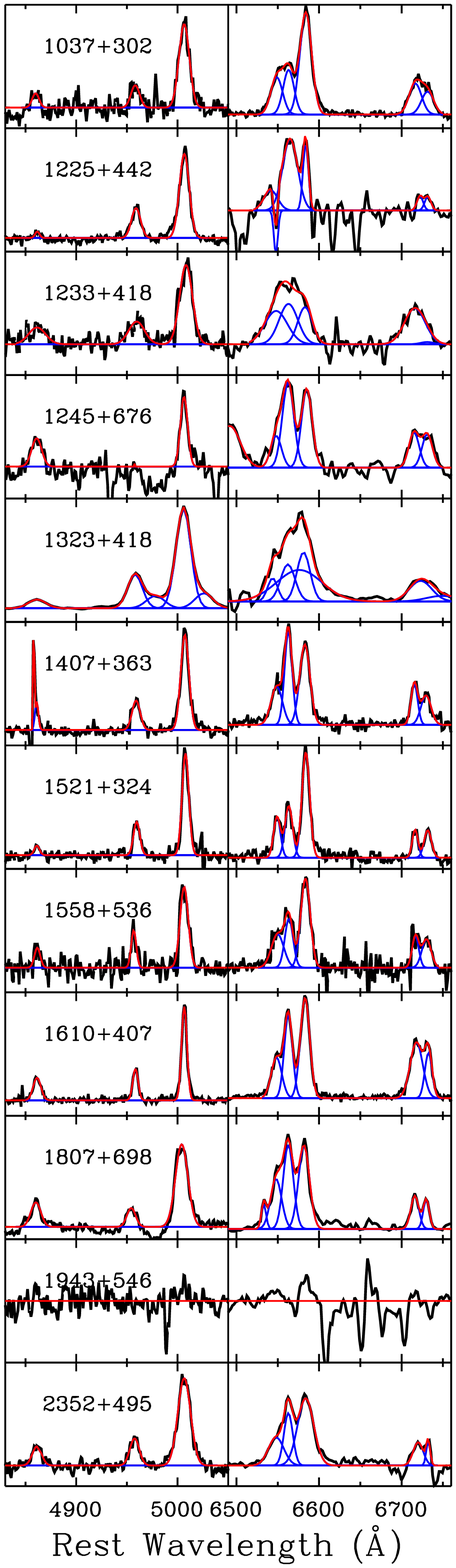} 
\caption{\Hb\ and \Ha\ regions of Type 2 AGNs. Narrow \Hb\ and \Ha\ lines, 
and \oiii, [N {\sc ii}] and [S {\sc ii}] in the continuum-subtracted spectra (black lines) 
were simultaneously fitted with Gaussian models (blue). 
The combined fit is represented by red lines.
(A color version of this figure is available in the online journal.)}
\label{specfit3}
\end{figure}

\begin{figure}
\center
    \includegraphics[width=0.236\textwidth, height=.8\textwidth]{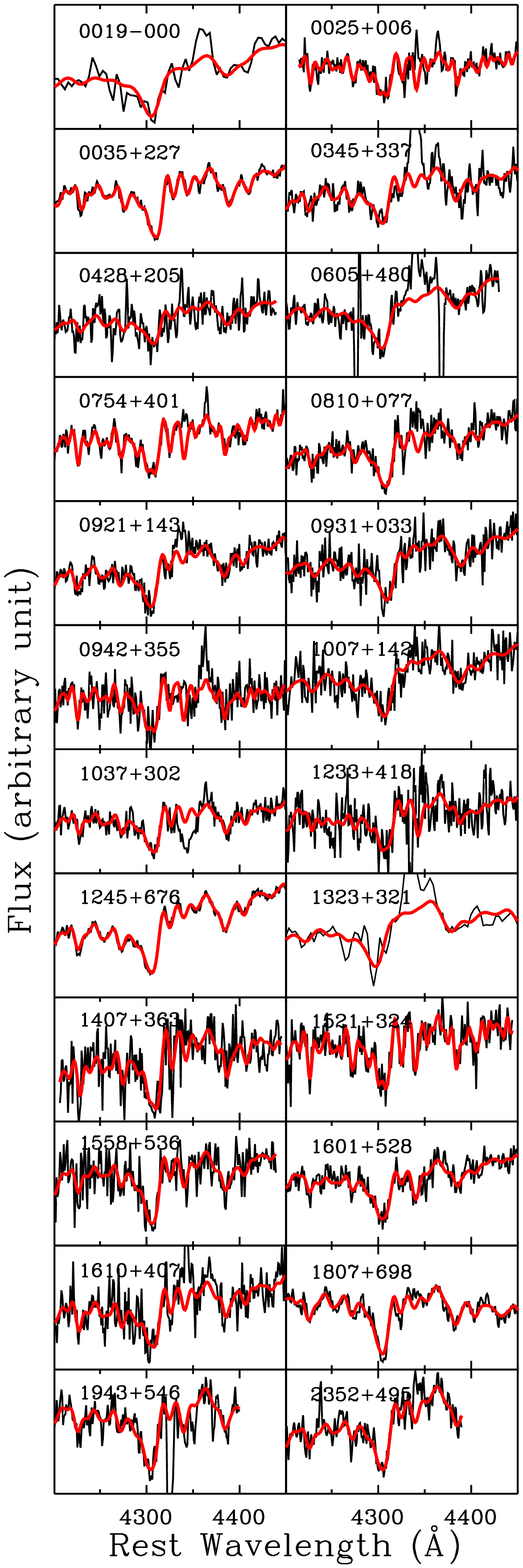} 
    \includegraphics[width=0.236\textwidth, height=.8\textwidth]{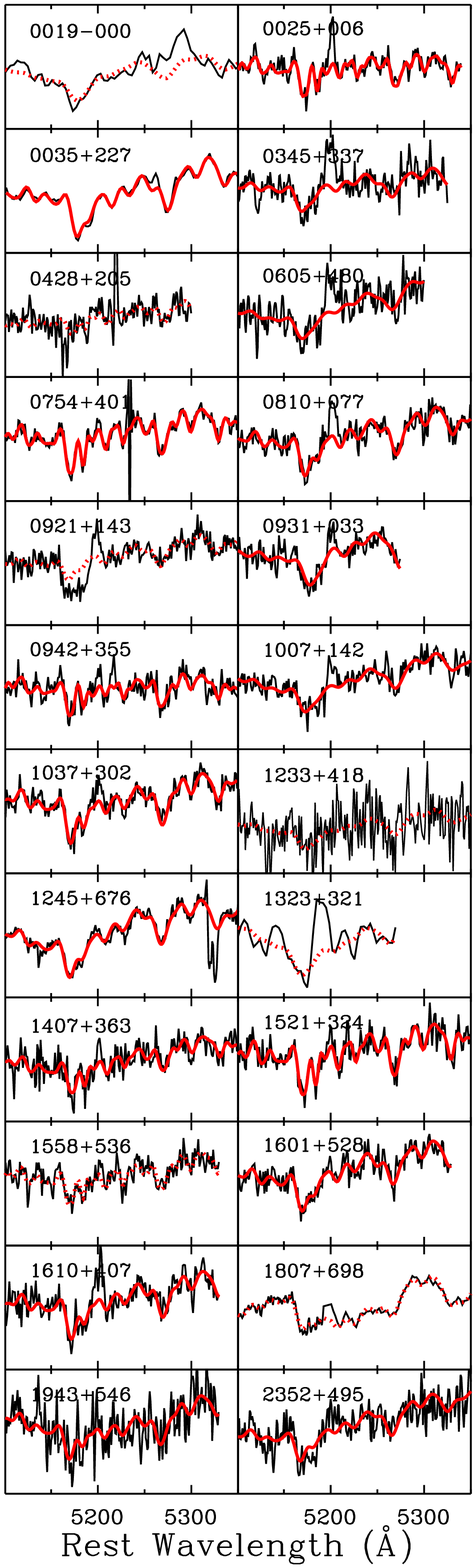} \\
\caption{Stellar velocity dispersion fitting in the G band (4300 \AA) ({\it left}), Mg $b$ 
triplet region ({\it right}).
The observed spectra (black lines) were compared with the best stellar template models (red lines). When the quality of the fit is poor (thin dashed lines), 
the measurements are discarded.
Note that non-stellar lines, e.g., H$\gamma$, [O {\sc iii}]$\lambda$4363, 
and [N {\sc i}]$\lambda$5200 lines were masked out before the fitting procedure. 
(A color version of this figure is available in the online journal.)}
\label{svdfit}
\end{figure}


\begin{deluxetable*}{lccccccccccccc}                                                                                                                                              
\tablewidth{0pt}                                                                                                                                                                  
\tabletypesize{\scriptsize}                                                                                                                                                       
\tablecaption{Narrow emission-line fluxes for the line diagnostics}                                                                                                               
\tablehead{                                                                                                                                                                       
    \colhead{Name} & \colhead{[O {\sc ii}]} & \colhead{[Ne {\sc iii}]} & \colhead{[O {\sc iii}]} & \colhead{H$\beta$}                                                                         
   & \colhead{[O {\sc iii}]} & \colhead{[O {\sc iii}]}& \colhead{[O {\sc i}]} & \colhead{H$\alpha$}                                                                                           
   & \colhead{[N {\sc ii}]} & \colhead{[S {\sc ii}]}    & \colhead{[S {\sc ii}]} & \colhead{[Ar {\sc iii}]}  & \colhead{E.I.}\\                                                                     
 \colhead{}  & \colhead{3727} & \colhead{3869} & \colhead{4363}                                                                                                             
  & \colhead{4861} & \colhead{4959} & \colhead{5007} & \colhead{6300} & \colhead{6563}                                                                                      
  & \colhead{6583} & \colhead{6716}   & \colhead{6731} & \colhead{7136} & \colhead{}    \\                                                                                  
   \colhead{(1)} & \colhead{(2)} &  \colhead{(3)} &  \colhead{(4)} &  \colhead{(5)}                                                                                         
  &  \colhead{(6)} &  \colhead{(7)} &  \colhead{(8)} &  \colhead{(9)} &  \colhead{(10)}                                                                                     
  &  \colhead{(11)} & \colhead{(12)} &  \colhead{(13)} & \colhead{(14)}                                                                                                     
 }                                                                                                                                                                          
 \startdata                                                                                                                                                                 
0019$-$000	&  4.7   		&  0.9   	  & \nodata   & 0.8     	& 0.8     		& 6.6    	& \nodata	& 6.0   		& 6.5     & 3.8	  		& 1.4     		& \nodata   		& \nodata	\\  
0025+006  	&  159.7 		&  25.2  		& \nodata   & 19.6  		& 47.1    		& 143.3  	& 25.8   	& 96.9  		& 151.9  	& 57.6			& 47.4    		& 1.8   				& 0.98   	\\  
0035+227  	&  110.8 		&  \nodata	& \nodata   & 14.0  		& \nodata  		& 51.4   	& \nodata	& 66.0  		& 50.8   	& \nodata 	& \nodata 		& \nodata   		& \nodata	\\  
0134+329  	&  100.4  	&  72.8 		& \nodata   & 104.2 		& 198.2   		& 615.2  	& 28.2   	& 372.9 		& 110.7  	& \multicolumn{2}{c}{43.2$\dag$}& \nodata & 1.63     \\ 
0221+276  	&  33.9  		&  11.8  		& 5.4       & 15.9  		& 44.4    		& 131.7  	& 9.0    	& 51.8  		& 26.7   	& 8.7	  		& 12.8    		& \nodata   		& 1.39   	\\  
0316+413  	&  3395.6 	&  1201.0		& \nodata   & 682.2 		& 2467.0  		& 7594.1 	& 3074.4 	& 3377.9		& 4160.5 	& 1894.2		& 1988.2  		& 65.1    			& 1.01   	\\  
0345+337  	&  140.9  	&  21.8  		& \nodata   & 26.2  		& 58.4    		& 178.4  	& 27.2   	& 83.9  		& 140.1  	& 60.0	  	& 38.1    		& 5.5     			& 0.90   	\\  
0428+205  	&  25.0   	&  \nodata	& \nodata   & 3.4   		& 6.3     		& 20.5   	& 26.6   	& 39.4  		& 53.5   	& 18.4	  	& 19.4    		& \nodata   		& 0.80   	\\  
0605+480  	&  53.0   	&  6.3   		& \nodata   & 8.0   		& 11.2    		& 34.7   	& 10.7   	& 34.0  		& 38.1   	& 18.1	  	& 14.4    		& \nodata   		& 0.79   	\\  
0754+401  	&  76.3   	&  20.7  		& 2.1       & 20.9  		& 61.8    		& 187.7  	& 14.1   	& 108.2 		& 98.8   	& 34.6	  	& 29.5    		& 4.7     			& 1.34   	\\  
0810+077  	&  67.1   	&  6.0   		& \nodata   & 12.2  		& 9.5     		& 27.8   	& 22.7   	& 68.9  		& 118.5  	& 48.6	  	& 44.5    		& \nodata   		& 0.40   	\\  
0921+143  	&  79.0   	&  5.2   		& \nodata   & 13.9  		& 4.6     		& 13.1   	& 26.0   	& 41.7  		& 96.5   	& 35.5	  	& 44.0    		& \nodata   		& -0.17  	\\  
0931+033  	&  30.2   	&  3.2   		& \nodata   & 4.5   		& 4.4     		& 16.3   	& 7.0    	& 27.1  		& 34.6   	& 14.3	  	& 14.8    		& \nodata   		& 0.71   	\\  
0941$-$080	&  18.8   	&  3.1   		& \nodata   & 4.3   		& 4.5     		& 15.2   	& 4.6    	& 12.2  		& 11.8   	& 4.7	   		& 6.3     		& 6.0     			& 0.71   	\\  
0942+355  	&  22.3   	&  13.6  		& \nodata   & 11.0  		& 31.1    		& 93.3   	& \nodata	& 22.6  		& 21.5   	& 6.4	   		& 5.6     		& \nodata   		& \nodata	\\  
1007+142  	&  31.0   	&  \nodata	& \nodata   & 4.8   		& 2.6     		& 7.3    	& 7.7    	& 13.5  		& 35.8   	& 12.7	  	& 12.7    		& \nodata   		& 0.03   	\\  
1037+302  	&  72.4   	&  14.6  		& \nodata   & 3.0   		& 5.8     		& 26.3   	& 17.7   	& 53.4  		& 133.2  	& 42.8	  	& 31.0    		& \nodata   		& 0.92   	\\  
1154+435  	&  37.4   	&  23.4  		& 5.5       & 19.1  		& 66.5    		& 200.2  	& 5.1    	& 63.7  		& 30.0   	& 13.4  		& 11.8    		& \nodata   		& 1.63   	\\  
1203+645  	&  17.7   	&  9.6   		& 1.6       & 14.3  		& 41.4    		& 117.9  	& 11.9   	& 61.9  		& 52.1   	& \multicolumn{2}{c}{37.4$\dag$}& \nodata & 1.32   	\\  
1225+442  	&  6.4    	&  5.4   		& \nodata   & \nodata 	& 12.3    		& 37.5   	& 3.3    	& 39.7  		& 13.2   	& 3.2	   		& 3.2     		& \nodata  			& \nodata	\\  
1233+418  	&  52.7   	&  \nodata	& \nodata   & 11.4  		& 14.4    		& 40.8   	& 13.9   	& 37.2  		& 26.4   	& \multicolumn{2}{c}{38.1$\dag$}& \nodata & 0.74    	\\
1245+676  	&  34.4   	&  \nodata	& \nodata   & 7.6   		& \nodata  		& 13.8   	& \nodata	& 24.2  		& 20.8   	& 9.0	   		& 8.5     		& \nodata  			& \nodata	\\  
1250+568  	&  93.0   	&  35.9  		& 19.8      & 39.0  		& 95.1    		& 288.4  	& 12.7   	& 139.4 		& 32.7   	& 32.8	  	& 19.8    		& \nodata  			& 1.57   	\\  
1323+321  	&  19.2   	&  7.4   		& 0.9       & 7.0   		& 32.7    		& 79.0   	& 22.1   	& 19.6  		& 23.9   	& \multicolumn{2}{c}{17.0$\dag$}& \nodata & 1.03    	\\
1345+125  	&  131.4  	&  44.1  		& \nodata   & 18.0  		& 210.7   		& 522.0  	& 113.0  	& 158.1 		& 130.1  	& 57.0	  	& 39.4     		& 18.4	  & 1.54   	\\        
1404+286  	&  50.7   	&  30.5  		& \nodata   & 20.0  		& 68.7    		& 212.6  	& 22.5   	& 86.1  		& 88.9   	& 13.0	  	& 38.7    		& \nodata   		& 1.29   	\\  
1407+363  	&  18.3   	&  2.5   		& \nodata   & 3.8   		& 9.6     		& 31.6   	& 9.8    	& 26.0  		& 30.8   	& 10.1	  	& 11.1    		& \nodata   		& 1.07   	\\  
1521+324  	&  15.6   	&  6.2   		& \nodata   & 2.3   		& 9.6     		& 31.2   	& 4.2    	& 15.9   		& 33.4   	& 6.5	   		& 8.1     		& \nodata   		& 1.23   	\\  
1558+536  	&  27.7   	&  4.0   		& \nodata   & 2.9   		& 4.7     		& 16.6   	& 4.2    	& 12.4  		& 25.8   	& 6.4	   		& 8.1     		& \nodata   		& 0.79   	\\  
1601+528  	&  17.2   	&  \nodata	& \nodata   & 2.6   		& 2.1     		& 12.9   	& 7.9    	& 17.9  		& 29.1   	& 22.9	  	& 9.4     		& \nodata   		& 0.66   	\\  
1610+407  	&  76.8   	&  7.6   		& \nodata   & 14.0  		& 13.6    		& 41.0   	& 21.5   	& 59.0  		& 82.4   	& 60.4	  	& 28.3    		& \nodata   		& 0.51   	\\  
1807+698  	&  116.9  	&  40.7  		& \nodata   & 24.6  		& 16.5    		& 103.6  	& 79.2   	& 125.5 		& 135.2  	& 39.3	  	& 26.4    		& \nodata   		& 0.77   	\\  
1943+546  	&  1.3    	&  \nodata	& \nodata   & \nodata		& \nodata 		& \nodata	& \nodata	& \nodata 	& \nodata	& \nodata 	& \nodata 		& \nodata   		& \nodata	\\  
2352+495		&  15.5   	&  2.7   		& \nodata		& 3.2   		& 4.6     		& 20.2   	& 4.5    	& 10.3  		& 24.0   	& \multicolumn{2}{c}{5.2$\dag$}	& \nodata & 0.90        
\enddata                                                                                                                                                                          
\tablecomments{Columns: (1) Target name;  (2)--(13) The flux of each narrow emission line in units of $10^{-16}$ ergs s$^{-1}$ cm$^{-2}$. $\dag$ Sum of blended [S {\sc ii}]6716, 6731;                                      
(14) Excitation index value derived from Equation (4).}                                                                                                                                  
\label{flux}                                                                                                                                                                      
\end{deluxetable*}                                                                                                                                                                


\subsection{Emission Lines}{{\label{measure}}

We measured the flux of all emission lines in the rest-frame 3727-7136\AA\ range,
including  [O {\sc ii}]$\lambda$3727, H$\beta$, \oiii, [O {\sc i}]$\lambda$6300, 
H$\alpha$, [N {\sc ii}]$\lambda$6583, [S {\sc ii}]$\lambda$6716/6731, [Ar {\sc iii}]$\lambda$7136, 
in order to constrain physical properties of NLR.
We fitted each emission line with Gaussian profiles using the IDL routine {\it mpfit} (Markwardt 2008). 
The fitting code determined the best $\chi^{2}$ fit, and measures the peak 
intensity, central wavelength, and line dispersion ($\sigma$).
After subtracting a linear continuum, emission lines were
modeled with single- or multi-Gaussian components.
For the \Hb+\oiii\ and [N {\sc ii}]+\Ha+[S {\sc ii}] regions,
we simultaneously fitted individual lines accounting for line blending. 
The flux and uncertainty were evaluated by integrating the fitted Gaussian function 
and the propagation of errors, respectively. 

In the case of Type 1 AGNs, it was necessary to simultaneously fit broad and narrow 
components of the \Ha\ and \Hb\ emission lines, in order to properly measure the 
narrow line fluxes.
One Type 1 object, {\it 0134+329} showed strong \ion{Fe}{2} emission in the \Hb+\oiii\ 
region, thus, we subtracted the \ion{Fe}{2} emission features,
by fitting them with a series of \ion{Fe}{2} templates convolved with various Gaussian 
velocities as performed in our previous studies (Woo \etal~2006; McGill \etal~2008;
see also, Boroson \& Green 1992).
For Type 2 AGNs, stellar absorption lines were subtracted before emission 
line fitting (see their Section~\ref{svd-sec}).
In Figure~\ref{specfit3} we present emission-line model fits around the \Hb\ and \Ha\ regions.
The measured emission-line fluxes are listed in Table~\ref{flux}.

We compared the measured emission-line fluxes with the available values in the previous studies
and found that in most cases emission-line fluxes were consistent within the measurement uncertainties. 
In the case of {\it 0221+276} and {\it 1250+568}, the [O {\sc iii}]$\lambda$5007 fluxes 
were smaller than those of O'Dea \etal~(2002) by 60 and 93\%, respectively, and
the widths (FWHM) of \oiii\ were $\sim 1.5$ times wider 
than those of O'Dea \etal~(2002), presumably due to the 
uncertainties in spectroscopic flux calibrations and systematic differences 
in fitting analysis. 
We assume our measurements suffer less systematic uncertainties since more sophisticated
fitting procedures were performed including stellar absorption line subtraction and 
multi-component analysis.

\subsection{Stellar Velocity Dispersions}{\label{svd-sec}}

We were able to measure the stellar velocity dispersions ($\sigma_{\ast}$) 
of 24 YRGs including two Type 1 objects, by comparing strong stellar lines, e.g., 
the G band (4300\AA), Mg $b$ triplet (5172\AA), and \ion{Ca}{2} H \& K lines, with
stellar templates (e.g., Woo \etal~2004, 2005; Bennert \etal~2011).
For most of Type 1 objects, stellar lines are relatively weak or not clearly detected 
due to the higher AGN continuum luminosity than stellar luminosity.
To measure stellar velocity dispersion, we utilized a Python-based code {\it vdfit}, 
which employs Bayesian statistical estimation using Markov Chain Monte Carlo sample 
with a set of stellar templates from the INDO-US library comprised of G and K 
giant stars (see Suyu \etal~2010).
AGN emission lines, i.e., H$\gamma$+[O {\sc iii}]$\lambda$4363 and [N {\sc i}]$\lambda$5200 
were masked out before the fitting procedure.
The continuum was fitted with a low-order polynomial function,
and then the width of absorption lines in the normalized spectra was compared 
to a series of stellar templates convolved with various Gaussian velocities.
The measured stellar velocity dispersions were corrected for the instrumental
resolution by subtracting the instrumental resolution from the measured velocity
dispersion in quadrature (e.g., Barth \etal~2002; Woo \etal~2004).
  
Figure~\ref{svdfit} presents the best-fit models and the observed spectra.
In most cases, the stellar velocity dispersions measured from the G band and Mg $b$ 
triplet regions were consistent within the measurement errors (see Figure~\ref{svdfit}). 
Thus, we adopted the average of the two measurements as a final value.
For several objects, e.g., {\it 0019+000}, {\it 1233+418}, {\it 1323+321}, 
{\it 1558+536}, {\it 1807+698} and {\it 1943+546},   
the Mg $b$ triplet region is not acceptable for measuring stellar velocity dispersion
due to the contamination of AGN emission lines and/or low S/N ratios.
Thus, we adopted velocity dispersion measured from the G-band region.
In the case of {\it 1225+442}, both G band and Mg $b$ regions have low S/N,
thus we measured $\sigma_{\ast}$  from the \ion{Ca}{2} H \& K region.
The measured stellar velocity dispersions will be used to estimate BH masses for each galaxy in next section.


\section{RESULTS}{\label{result}}

In this section, we investigate the physical properties of YRGs by 
estimating BH masses in Section~4.1, explore the properties of the NLR and 
the accretion rate in Section~4.2, and compare radio and emission-line properties in Section~4.3.

\begin{figure*}
\epsscale{1.18}\plotone{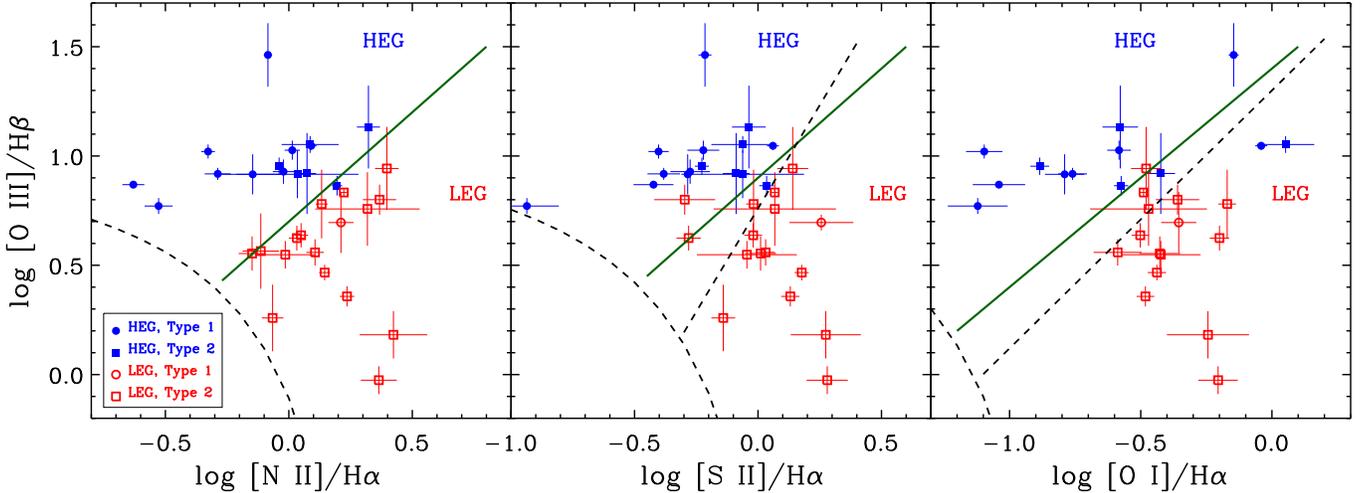}
\caption{Emission-line flux ratios. Seyfert, LINER, and star-forming galaxies classifications
are indicated by dashed lines following Kewley \etal~(2006). YRGs are classified into two
groups: HEGs (filled blue symbols) and LEGs (open red symbols), while Type 1 (Type 2) objects are denoted 
by circle (rectangle) symbols. The dividing lines between HEGs and LEGs are indicated
by green solid lines. 
(A color version of this figure is available in the online journal.)
}
\label{bpt}
\end{figure*}


\subsection{Black Hole Mass}

We determine BH masses (\mbh) using two different methods for Type 1 and Type 2 AGNs,
respectively. For 10 Type 1 AGNs with detected broad emission lines, we estimate \mbh\ 
based on the virial assumption of the broad-line region (e.g., Woo \& Urry 2002; McGill \etal~2008; Shen \etal~2008).
In practice, we used the single-epoch mass estimator from McGill \etal~(2008):
\begin{eqnarray}
\mbh/\msun &=& 10^{7.21}  \left(\frac{\sigma_{\rm H\alpha}}{10^{3}~\mathrm{\kms}}\right)^{2} \left(\frac{L_{\rm H\alpha}}{10^{42} ~\mathrm{erg~ s^{-1}}}\right)^{0.55},
\end{eqnarray}
where $\sigma_{\rm H\alpha}$ is the line dispersion of broad H$\alpha$,
and L$_{\rm H\alpha}$ is the broad \Ha\ luminosity. 
Table~\ref{Mbh} lists the determined \mbh\ along with the luminosity and line dispersion
of broad \Ha\ for 10 Type 1 AGNs.

Since there are several recipes of \mbh\ estimation based on different calibrations, 
we also estimated \mbh for comparison using the single-epoch mass estimator from 
Park \etal~(2012) with the virial coefficient determined from the \msigma\ relation 
of the reverberation sample (Woo \etal~2010): 
\begin{eqnarray}
\mbh/\msun &=& 10^{7.60}  \left(\frac{\sigma_{\mathrm H\beta}}{10^{3}~\mathrm{\kms}}\right)^{2} \left(\frac{\lambda L_{5100}}{10^{44}  ~\mathrm{erg~ s^{-1}}}\right)^{0.52},
\label{eq:L5100}
\end{eqnarray}
where $\sigma_{\mathrm H\beta}$ is the line dispersion of broad H$\beta$,
and L$_{\rm 5100}$ is AGN continuum luminosity at 5100 \AA. 
Since broad H$\beta$ is relatively weak and detected only for 7 objects,
we inferred $\sigma_{H\beta}$ from $\sigma_{H\alpha}$, using the correlation between 
H$\alpha$ and H$\beta$ line widths (Equation (3) in Greene \& Ho 2005).
In the case of L$_{5100}$, we used the correlation between L$_{\rm H\alpha}$ and 
L$_{5100}$ given by Greene \& Ho (2005).
\mbh\ estimated in this method is systematically larger than that determined
with Equation (1) by $\sim26\%$ due to the different calibrations between luminosities and/or 
line widths adopted in the mass estimators.
However, the systematic difference is relatively small compared to the large range
of mass and luminosity of the sample, and does not significantly affect the 
main results.

For objects with measured $\sigma_*$, we derive \mbh\ from $\sigma_{\ast}$ using 
the $\mbh-\sigma_{\ast}$ relation of early-type galaxies (G{\"u}ltekin \etal~2009):  
\begin{equation}
\log \mbh/\msun = 8.22 + 3.86 \log (\sigma_{\ast} / {200~\mathrm{\kms}}).
\label{eq:msigma}
\end{equation}
For two Type 1 AGN {\it 0942+355} and {\it 1601+528}, \mbh\ was determined 
with both methods, however, we adopted \mbh\ estimates from the \msigma\ relation
since the systematic uncertainties in Equation (1) and (2) due to dust extinction and 
indirect estimates of the broad-line size from \Ha\ luminosity are probably large
for these Type 1 AGNs with relatively redder spectral energy distributions (SEDs).
In summary, \mbh\ was determined from measured $\sigma_{\ast}$ for 24 Type 2 objects and 2 Type 1 objects 
as listed in Table~\ref{Mbh2}.
The \mbh\ of the YRGs in our sample ranges over two orders of magnitude, 
$7.0 < {\rm log (\mbh/\msun)} < 9.2$, indicating that YRGs host relatively
massive BHs, similar to the large-scale radio galaxies.

\setlength{\tabcolsep}{3.5pt}   

\begin{deluxetable}{lccccc}
\tablewidth{0pt}                                                                                                                           
\tabletypesize{\scriptsize}                                                                                                                
\tablecaption{\mbh\ and gas properties of Type 1 AGNs}
\tablehead{                                                                                                                                              
\colhead{Name} & \colhead{log $L_{H\alpha}$}  & \colhead{$\sigma_{H\alpha}$} & \colhead{log \mbh} & \colhead{log $T_{e}$}	& \colhead{ log	$n_{e}$ } \\   
\colhead{}     & \colhead{(erg s$^{-1}$)}     & \colhead{(\kms)}             & \colhead{(\msun)}  & \colhead{(K)}  				& \colhead{(cm$^{-3}$)} \\     
\colhead{(1)}  & \colhead{(2)}  							& \colhead{(3)} 							 & \colhead{(4)}      & \colhead{(5)}          & \colhead{(6)}               
}                                                                                                                                                        
\startdata                                                                                                                                               
0134+329    & 43.8			& 1549 						& 8.6			& \nodata	 & \nodata		\\          	                                                               
0221+276    & 42.7			& 3531						& 8.7			& 4.4			 & 3.4				\\          	                                                               
0316+413    & 41.6			& 1032						& 7.0			& \nodata	 & 2.9				\\          	                                                               
0942+355    & 42.2			& 1323						& 7.5    	& \nodata  & 2.5				\\   	                                                                       
1154+435    & 42.8			& 1268 						& 7.9			& 4.3			 & 2.5				\\          	                                                               
1203+645    & 42.8			& 2567    				& 8.4   	& 4.1			 & \nodata		\\          	                                                               
1250+568    & 43.2			& 1705						& 8.3			& 4.6			 & \nodata		\\          	                                                               
1345+125    & 41.8			& 1307						& 7.3			& \nodata	 & \nodata		\\          	                                                               
1404+286    & 42.6			& 3990						& 8.7			& \nodata	 & \nodata		\\          	                                                               
1601+528    & 41.5			& 1139						& 7.1			& \nodata  & \nodata		   	                                                                       
\enddata                                                                                                                                                 
\tablecomments{Columns: (1) Target name;                                                                                                                 
(2) Luminosity of the broad H$\alpha$;
(3) Line dispersion of the broad H$\alpha$;
(4) \mbh\ estimated with single epoch method;
(5) Electron temperature estimated from [O {\sc iii}] ratio (I(4959+5007)/I(4363)),
assuming electron density of $n_{e} \sim 10^3$ cm$^{-3}$;                                                                                             
(6) Electron  density  estimated from  the [S {\sc ii}] ratio of I(6716)/I(6731),
assuming an electron temperature of $10^4$ K.  }                                                                                                         
\label{Mbh}                                                                                                                                              
\end{deluxetable}                           

\begin{deluxetable}{llccc}
\tablewidth{0pt}                                                                                                                           
\tabletypesize{\scriptsize}                                                                                                                
\tablecaption{\mbh\ and gas properties of Type 2 AGNs}
\tablehead{                                             
\colhead{Name} & \colhead{$\sigma_\ast$} & \colhead{log \mbh} & \colhead{log $T_{e}$}	& \colhead{ log	$n_{e}$ } \\
\colhead{}     & \colhead{(\kms)}             & \colhead{(\msun)}  & \colhead{(K)}  				& \colhead{(cm$^{-3}$)} \\     
\colhead{(1)}  & \colhead{(2)}  							& \colhead{(3)} 							 & \colhead{(4)}      & \colhead{(5)}         
}                                                                                                                                                        
\startdata                                                                                                                                               
0019$-$000      & 329 $\pm$ 114 	& 9.1    	& \nodata	 & \nodata		\\   	                                                                       
0025+006        & 108 $\pm$ 10 	  & 7.2    	& \nodata	 & 2.3				\\   	                                                                       
0035+227        & 222 $\pm$ 7  	  & 8.4    	& \nodata	 & \nodata		\\   	                                                                       
0345+337        & 258 $\pm$ 26 	  & 8.6    	& \nodata	 & \nodata		\\   	                                                                       
0428+205        & 263 $\pm$ 26 	  & 8.7    	& \nodata	 & 2.9				\\   	                                                                       
0605+480        & 360 $\pm$ 42 	  & 9.2    	& \nodata	 & 2.2				\\   	                                                                       
0754+401        & 122 $\pm$ 7  	  & 7.4    	& 4.1	 		 & 2.5				\\   	                                                                       
0810+077        & 233 $\pm$ 14 	  & 8.5    	& \nodata	 & 2.6				\\   	                                                                       
0921+143        & 248 $\pm$ 19 	  & 8.6    	& \nodata	 & 3.1				\\   	                                                                       
0931+033        & 341 $\pm$ 24 	  & 9.1    	& \nodata	 & 2.8				\\   	                                                                       
0941$-$080      &  148 $\pm$ 17$^\ddag$   & 7.7	 		& \nodata	 & 3.2				\\   	                                                                       
0942+355$^\dag$    		& 120 $\pm$ 17		& 7.4	  	& \nodata	 & \nodata		\\   	                                                                       
1007+142        & 344 $\pm$ 43 	  & 9.1    	& \nodata	 & 2.8				\\   	                                                                       
1037+302	      & 199 $\pm$ 14 	  & 8.2    	& \nodata	 & 1.6				\\	                                                                         
1225+442        & 184 $\pm$ 59 	  & 8.1    	& \nodata	 & 2.8				\\   	                                                                       
1233+418        & 166 $\pm$ 22 	  & 7.9    	& \nodata	 & \nodata		\\   	                                                                       
1245+676        & 233 $\pm$ 2  	  & 8.5    	& \nodata	 & 2.7				\\   	                                                                       
1323+321        & 353 $\pm$ 79 	  & 9.2    	& 4.1	 		 & \nodata		\\   	                                                                       
1407+363        & 154 $\pm$ 19 	  & 7.8    	& \nodata	 & 2.9				\\   	                                                                       
1521+324        & 118 $\pm$ 11 	  & 7.3    	& \nodata	 & 3.1				\\   	                                                                       
1558+536        & 229 $\pm$ 30 	  & 8.4    	& \nodata	 & 3.2				\\   	                                                                       
1601+528$^\dag$    		& 240 $\pm$ 15  	& 8.5	  	& \nodata	 & \nodata	\\	                                                              
1610+407        & 201 $\pm$ 24 	  & 8.2    	& \nodata	 & \nodata		\\   	                                                                       
1807+698        & 258 $\pm$ 21 	  & 8.6    	& \nodata	 & \nodata		\\     	                                                                     
1943+546        & 234 $\pm$ 28    & 8.5     & \nodata  & \nodata		\\                                                                           
2352+495        & 246 $\pm$ 14    & 8.6     & \nodata  & \nodata	   	                                                                             
\enddata                                                                                                                                                 
\tablecomments{Columns: (1) Target name.                                                                                                                 
$^\dag$Type 1 AGN; 
(2) Stellar velocity dispersion.                                                                                                                          
$^\ddag$We adopted the measurement from Snellen \etal~(2003).;
(3)  \mbh\ estimated from the $\mbh-\sigma_{\ast}$ relation:  
(4) Electron temperature estimated from [O {\sc iii}] ratio (I(4959+5007)/I(4363)),
assuming electron density of $n_{e} \sim 10^3$ cm$^{-3}$;                                                                                             
(5) Electron  density  estimated from  the [S {\sc ii}] ratio of I(6716)/I(6731),
assuming an electron temperature of $10^4$ K.  }                                                                                                         
\label{Mbh2}                                                                                                                                              
\end{deluxetable}                           

\setlength{\tabcolsep}{5pt}   
 

\begin{figure}
\center
\epsscale{1}\plotone{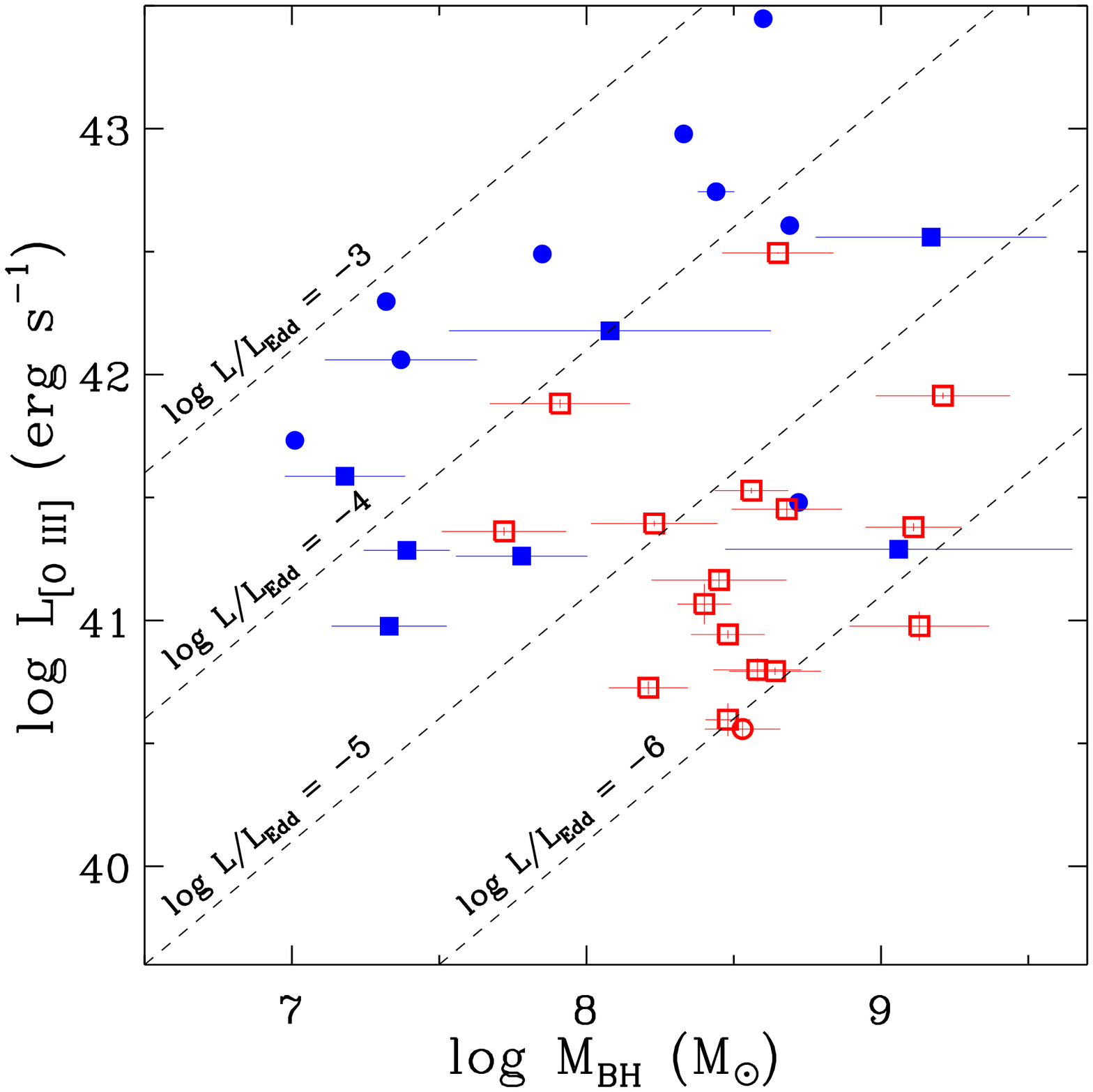}
\caption{Comparison of the \oiii\ line luminosity with \mbh. 
Symbols are the same as in Figure~\ref{bpt}. 
Dashed lines indicate the ratio of line luminosity to the Eddington limit. 
(A color version of this figure is available in the online journal.)
}
\label{Mbh-Lo3}
\end{figure}

\begin{figure}
\center
\epsscale{1}\plotone{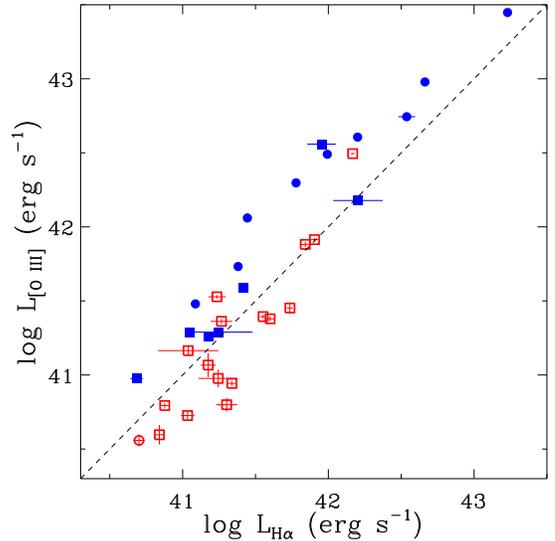}
\caption{\Ha\ vs. \oiii\ line luminosities. Note that HEGs (filled symbols) have higher 
\oiii\ luminosity than LEGs (open symbols) at fixed \Ha\ luminosity as expected from higher \oiii/\Hb\ ratio. 
(A color version of this figure is available in the online journal.)}
\label{Lo3ha}
\end{figure}

\begin{figure}
\center
\epsscale{1}\plotone{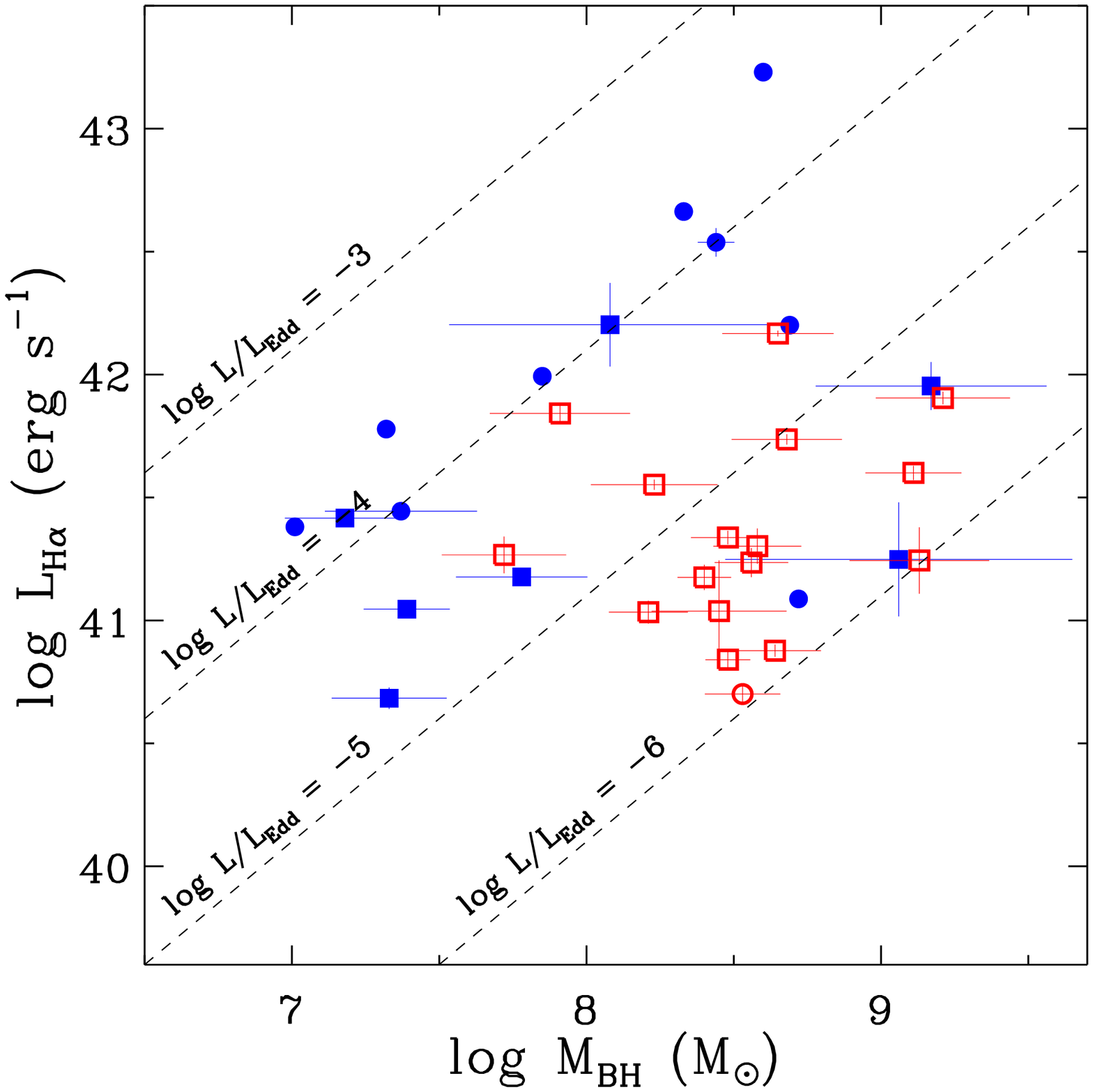}
\caption{Comparison of the \Ha\ line luminosity with \mbh. 
Symbols are the same as in Figure~\ref{bpt}. 
Dashed lines indicate the ratio of line luminosity to the Eddington limit. 
(A color version of this figure is available in the online journal.)
}
\label{Mbh-LHa}
\end{figure}

\begin{figure}
\center
\epsscale{1.}\plotone{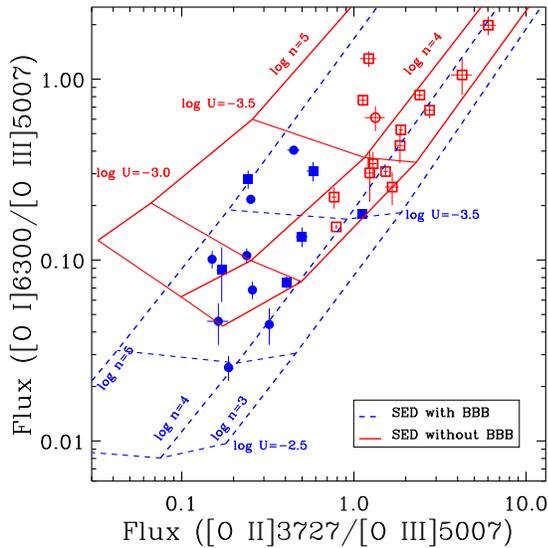}
\caption{[O {\sc i}]$\lambda$6300/[O {\sc iii}]$\lambda$5007 vs. [O {\sc ii}]$\lambda$3727/[O {\sc iii}]$\lambda$5007 ratios. The measured line ratios of HEGs (filled symbols) and LEGs (open symbols) are compared
with photoionization model predictions with a SED with BBB (blue lines) and a SED without BBB
(red lines), respectively. The assumed ionization parameters (log U) and hydrogen densities (log n)
are denoted by different lines.
(A color version of this figure is available in the online journal.)
}
\label{o1o3}
\end{figure}

\subsection{Narrow-line Region}

\subsubsection{Excitation and Accretion}

To investigate the properties of accretion activities in YRGs, 
we compare the flux ratios of high and low excitation lines in Figure~\ref{bpt}.
All YRGs in our sample are classified as AGNs based on the criteria 
of Kewley \etal~(2006), suggesting that photoionization by a nuclear source
is a main mechanism of narrow line emission in YRGs. There is a wide
range of line flux ratios, which are generally divided
into Seyfert and LINER (Low-Ionization Nuclear Emission-line Region) classes 
in radio-quiet (RQ) Type 2 AGNs as indicated by diagonal dashed lines.

We classify YRGs into two groups: HEGs and LEGs based on the flux ratios
between high- and low-excitation lines using the excitation index (EI) 
suggested by Buttiglione \etal~(2010), 
\begin{eqnarray}
\scriptsize
{\rm EI} = \log {\rm [O~III] \over H\beta}  - {1 \over 3} \left(\log {\rm [N~II] \over H\alpha} + \log {\rm [S~II] \over H\alpha } + \log { \rm [O~I] \over H\alpha} \right), \nonumber\\
\end{eqnarray}
which represents the average ratio of high to low excitation line fluxes.
If EI is larger (smaller) than 0.95, YRGs are classified as HEG (LEG). 
For five objects ({\it 0019-000, 0035+227, 0942+355, 1225+442}, and {\it 1245+676}), 
not all three high-excitation lines (i.e., [\ion{N}{2}], [\ion{S}{2}] and [\ion{O}{1}]) 
were measured, thus, we used only one or two available lines for classification
as suggested by Buttiglione \etal~(2010). 

{\it 1943+546} is excluded since none of the three low-ionization lines
were detected.
We note that {\it 0345+337} is classified as LEG, while it was previously 
classified as HEG by Buttiglione \etal~(2010), presumably owing to the lower quality
of their spectra.   

In summary, the YRG sample consists of 16 HEGs, 17 LEGs, and 1 unclassified object.
All Type 1 AGNs with broad \Ha\ belong to the HEG class except for {\it 1601+528}, 
while Type 2 AGNs belong to both HEG and LEG classes.
These trends are similar to those found in FR I and FR II galaxies. For example,
Buttiglione \etal~(2010) showed that all Type 1 objects in their 3CR sample were HEGs, 
while Type 2 objects were composed of both HEGs and LEGs.

\subsubsection{Accretion Rate vs. High-to-low Excitation Line Ratio}

We investigate whether HEGs and LEGs have systematically different accretion 
activity by comparing their Eddington ratios.
Since bolometric luminosity cannot be directly measured for Type 2 AGNs, 
first, we use the \oiii\ line
luminosity as a proxy for bolometric luminosity.
In Figure~\ref{Mbh-Lo3} we compare the \oiii\ luminosity with \mbh. 
At fixed \mbh, HEGs have higher line luminosities 
than LEGs, suggesting that HEGs have higher Eddington ratios. 
Although, the separation between HEGs and LEGs is not a clear cut,
the average Eddington ratio of HEGs is $\sim$1.2 dex larger than that of LEGs.
 
However, \oiii\ may not be a good indicator of bolometric luminosity,
particularly when HEGs and LEGs are compared since HEGs have relatively 
higher \oiii/\Hb\ ratio than LEGs. Thus, bolometric luminosity based 
on \oiii\ can be systematically overestimated for HEGs. 
Also, an orientation dependency of \oiii\ has been reported as that for given isotropic flux 
(e.g., far-infrared or [OIV] 25.9\micron) the \oiii\ line flux of Type 2 AGNs is systematically
lower than that of Type 1 AGNs 
(Jackson \& Browne 1990; Nagao et al. 2001; Haas et al. 2005; Baum et al. 2010), 
suggesting a bias of \oiii\ as a bolometric luminosity indicator.
A recent study by
Netzer (2009) reported that bolometric correction of \oiii\  
systematically changes as a function of \oiii/\Hb\ ratio, while \Ha\ does
not show such a trend, indicating that \Ha\ is a better tracer of bolometric
luminosity.

To test the systematic difference between \oiii\ and \Ha,
we compare them in Figure~\ref{Lo3ha}.
As expected, HEGs have higher \oiii\ luminosity for given \Ha\ luminosity, 
while LEGs have lower \oiii\ luminosity for given \Ha\ luminosity.
This result indicates that if the luminosity of \oiii\ is used as a proxy
for AGN bolometric luminosity, the difference of bolometric luminosity
between HEGs and LEGs would be overestimated.

In Figure~\ref{Mbh-LHa} we compare \mbh\ with the \Ha\ luminosity.
The average difference between HEGs and LEGs decreases compared to 
Figure~\ref{Mbh-Lo3}. However, we find that the average Eddington ratio 
of HEGs is still larger than that of LEGs by $\sim$1.0 dex (a factor of 9.0).
The difference of the Eddington ratio suggests that HEGs may 
have higher mass accretion rate at fixed \mbh\ than LEGs or that 
radiative efficiency is systematically different 
if the mass accretion rate normalized by \mbh\ is similar. 

\subsubsection{Comparison with Photoionization Models}

We investigate whether HEGs and LEGs have different types of
photoionizing continua by comparing observed emission-line flux ratios with 
photoionization models. Using Cloudy version 08.00 (Ferland \etal~1998), 
we calculate the ratios of various emission lines arising in the NLR gas clouds. 
We assume the NLR metallicity $Z_{\rm NLR} = 2 Z_{\sun}$ since such a super-solar 
metallicity has been generally reported for the NLRs (e.g., Nagao \etal~2006),   
and investigate line flux ratios with a density range of $n_{\rm H} = 10^{2.0} - 10^{5.0}$ 
and an ionization parameter range of $\log U = 10^{-4.0} - 10^{-2.5}$ that 
are typical for NLRs (e.g., Nagao \etal~2001). 
In the case of photoionizing sources, we used two types of SEDs: SED with big 
blue bump (BBB) and SED without BBB (see Nagao \etal~2002; Kawakatu \etal~2009).
Note that the SED without BBB can be expressed with a single power-law continuum 
(harder spectrum) in the range of $\sim$10$^{12}$ to $\sim$10$^{20}$ Hz.
Such a harder spectrum of the optically thin disk (radiatively inefficient accretion
flow) generates different line ratio compared to the SED with BBB, as presented in Figure 10.
At fixed $U$ and $n_{\rm H}$, the photoionization models using SED with BBB (blue lines) 
predict lower [O {\sc i}]$\lambda$6300/[O {\sc iii}]$\lambda$5007
than the models using SED without BBB (red lines),
as also summarized by Kawakatu \etal~(2009; see Section 4.1).

We compare the measured line flux ratios of 
[O {\sc i}]$\lambda$6300/[O {\sc iii}]$\lambda$5007 and 
[O {\sc ii}]$\lambda$3727/[O {\sc iii}]$\lambda$5007 with model predictions 
in Figure~\ref{o1o3}, where HEGs and LEGs are located in different regions.
For HEGs, [O {\sc iii}]$\lambda$5007 is stronger than [O {\sc i}]$\lambda$6300 or [O {\sc ii}]$\lambda$3727 lines, consistent with the photoionization model calculations
using SEDs with BBB. In contrast, for LEGs [O {\sc iii}]$\lambda$5007 is relatively 
weak compared to [O {\sc i}]$\lambda$6300 or [O {\sc ii}]$\lambda$3727 lines,
suggesting that the measured line flux ratios of LEGs are consistent with
models predictions using SEDs without BBB.
We may interpret these findings as that HEGs generally have radiatively efficient
accretion disk with BBB, while LEGs have radiatively inefficient disk without BBB.
In other words, high and low excitation can be related with the properties
of accretion disk. 

In the case of SDSS Seyfert 2 galaxies, [O {\sc iii}]$\lambda$5007 is relatively
stronger than [O {\sc i}]$\lambda$6300 or [O {\sc ii}]$\lambda$3727, which
is consistent with SED with BBB (see Kawakatu \etal~2009). Thus, high-excitation
YRGs and Seyfert 2 galaxies may have similar accretion properties. 
Kawakatu \etal~(2009) suggested that YRGs have radiatively inefficient accretion
disk based on the small sample of YRGs, of which the oxygen line flux ratios 
were consistent with photoionization model without BBB. 
However, note that their sample was mostly composed of LEGs. In contrast, in this
work, we include many Type 1 AGNs with relatively higher Eddington ratios, 
which are mostly HEGs, 
leading to a more general view on the accretion properties of YRGs.

The different location on the oxygen line ratio diagram can also be
interpreted as that LEGs have relatively lower ionization parameter than HEGs, 
while both classes have similar ionizing SEDs.
We cannot rule out this possibility with the current data. To break the degeneracy
between the SED shape and the ionization parameter, we attempted to measure 
[Ar {\sc iii}]/[O {\sc iii}] ratio, which is a robust indicator of the ionization parameter, 
independent of SED shapes (Nagao et al. 2002).
Unfortunately, [Ar {\sc iii}] is generally very weak and we are not able to detect 
[Ar {\sc iii}] in most of LEGs, leading to inconclusive results (see Table~\ref{flux}).

\subsubsection{Gas Properties}

We estimated the electron temperature ($T_e$) and density ($n_e$) of the NLR using emission-line 
ratios.  The \oiii\ flux ratio,  I(4363)/I(4959+5007) is generally used as an 
indicator of $T_e$, while the flux ratio of [S {\sc ii}], I(6716)/I(6731) is 
used for estimating $n_e$ (Osterbrock \& Ferland 2006). 
We used the {\it temden} task in the IRAF {\it nebular} package to calculate $T_e$ and $n_e$.
For six objects we were able to estimate $T_e$ in the ranges of
 $4.0 \leq \log T_{e} \leq 4.6$, assuming $n_e = 10^3$ cm$^{-3}$. 

For measuring electron density, [S {\sc ii}] ratios were determined for 20 objects, ranging from $0.7$ to $1.4$, which corresponded to $\log n_e \approx 1.6 - 3.4$ cm$^{-3}$, 
assuming $n_e = 10^4$ K. 
For the remaining galaxies, the electron density could not be calculated due to the
failure of deblending of [S {\sc ii}]$\lambda$6717 \& [S {\sc ii}]$\lambda$6731 lines,
We find that electron temperature and density of the NLRs in YRGs are similar to
typical RQ AGNs. The estimated electron temperature and density of NLR are listed in 
Table~\ref{Mbh} and \ref{Mbh2}. 


\begin{figure}
\center
\epsscale{1.}\plotone{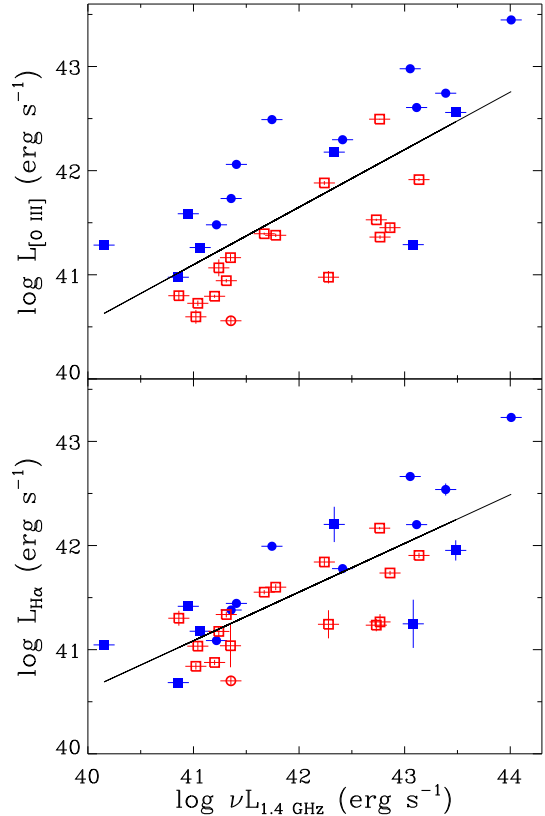}
\caption{Comparison of radio core luminosity with the \oiii\ (top) and \Ha\ line luminosities (bottom).
Symbols are the same as in Figure~\ref{bpt}. 
The measurements errors of radio luminosities are assumed to 0.1 dex. 
(A color version of this figure is available in the online journal.)
}
\label{Lradio}
\end{figure}

\begin{figure}
\center
\epsscale{1.}\plotone{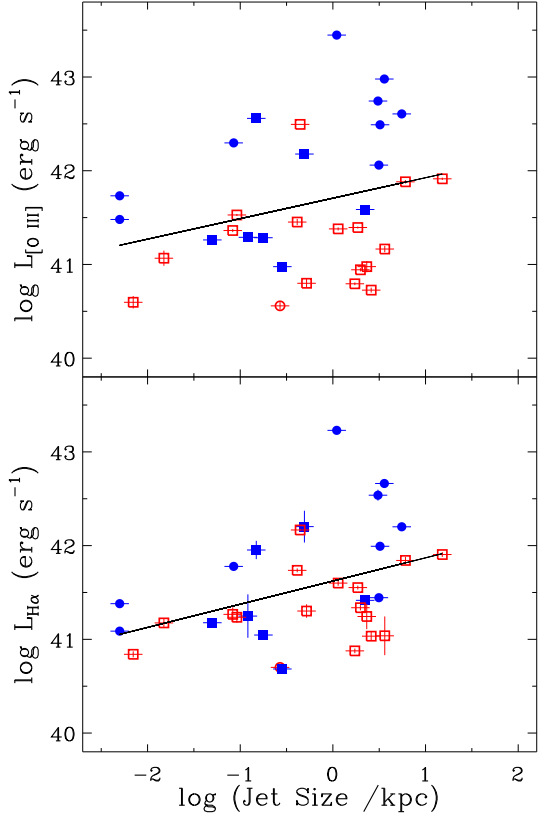}
\caption{Comparison of the jet size with the \oiii\ (top) and \Ha\ line luminosities (bottom).
Symbols are the same as in Figure~\ref{bpt}.
(A color version of this figure is available in the online journal.)
}
\label{JetSize}
\end{figure}

\subsection{Radio jet properties}{\label{Radio}}

We investigate the connection between radio and accretion activities
by comparing radio luminosity and jet size with \oiii\ and \Ha\ luminosities.
Figure~\ref{Lradio} compares the radio core luminosity with the line luminosities
of \oiii\ and \Ha. 
In general there is a broad correlation between radio and narrow emission-line luminosities.
If we consider narrow-line luminosity as a proxy for AGN bolometric luminosity as discussed in Section~4.2, 
the broad correlation indicates that radio and accretion activities are fundamentally
connected in YRGs. A similar trend has been reported in large-scale radio galaxies 
(i.e., FR Is and FR IIs) by a number of previous studies 
(e.g. Baum \& Heckman 1989; Rawlings \& Saunders 1991; Buttiglione \etal~2010). 

Comparing HEGs and LEGs at fixed radio luminosity, 
we find HEGs have higher \oiii\ luminosity than LEGs, suggesting that accretion
properties are different between HEGs and LEGs although radio activities are similar.
It is possible that there is a second parameter to change the accretion power at given
radio power. These trends were also reported for large-scale radio galaxies
by Buttiglione \etal~(2010), who investigated the radio and emission-line luminosities
of a sample of low-z 3C radio galaxies (see also Kunert-Bajaraszewska \& Labiano 2010). 
They argued that the separation between HEGs and LEGs is due to the different temperature of
the accreting gas. While HEGs have inflow of cold gas, LEGs accrete hot gas, 
resulting in a harder photoionizing spectrum and stronger low-excitation lines 
than those of HEGs.

 If the \oiii\ line flux is systematically higher than the \Ha\ line flux
in HEGs as discussed in Section~4.2 (see Figure~\ref{Lo3ha}), then the difference between HEGs and LEGs
may be overestimated. To overcome this systematic uncertainty, we also use the \Ha\ luminosity 
as a proxy for bolometric luminosity for investigating whether HEGs and LEGs have different 
accretion properties at fixed radio luminosity. 
When \oiii\ is replaced by \Ha\ (bottom panel of Figure~\ref{Lradio}), the separation 
between HEGs and LEGs is less clear although on average HEGs have higher \Ha\ luminosity
than LEGs. Since the division between HEGs and LEGs is not distinct, 
we derive the correlation for the combined sample of HEGs and LEGs 
using a least square fitting method:
\begin{eqnarray}
&& \log \rm L_{H\alpha} = (21.9 \pm 3.0) + (0.5 \pm 0.1) \log \nu \rm L_{1.4}. 
\end{eqnarray}
The correlation has 0.4 dex scatter, indicating that accretion luminosity can vary by more than
a factor of two at given radio luminosity. 
The correlation between narrow-line and radio luminosities found in YRGs seems similar to
that of large-scale radio galaxies. However, the slope of the correlation in large-scale
radio galaxies is close to 1, which somewhat steeper than that of YRGs (Buttiglione et al. 2010). 

By comparing the linear jet size with the \oiii\ and \Ha\ luminosities in Figure~\ref{JetSize}, 
we find a weak correlation between emission-line luminosity and the projected 
jet size, as similarly reported by Labiano (2008) for GPS and CSS sources.
However, the correlation is not very tight with considerably large scatter
in the case of \oiii.
When the \Ha\ luminosity is compared with the jet size, the relation becomes 
slightly tighter 
with 0.5 dex scatter, probably due to the systematic difference of \oiii/\Ha\ ratios
between HEGs and LEGs.
We derive the correlation between \Ha\ luminosity and jet size as:
\begin{eqnarray}
&& \log \rm L_{H\alpha} = (41.6 \pm 0.1) + (0.2 \pm 0.1) \log (R_{jet} /kpc), 
\end{eqnarray}
where R$_{\rm jet}$ is the size of jet in kpc.
Note that since the jet size is measured as a projected size, the true jet size
may be larger, implying that the correlation is even weaker.
The shallow slope of the relation indicates that YRGs with similar emission
line luminosities and accretion rates can have dramatically different jet sizes,
implying that the jet size may be determined by other mechanisms, e.g., the properties
of interstellar medium in the host galaxies than accretion properties.
We further investigate whether the HEG/LEG ratio changes as a function of the jet size
using our YRG sample and 3CR radio galaxies from Buttiglione \etal~(2010).
No significant change of the ratio has been detected over the large range of the 
jet size ($\sim$10 pc to $\sim$1 Mpc), suggesting that the jet size is not directly 
connected to the properties of the accretion flows. These findings are consistent 
with a scenario that accretion properties can change over the lifetime of radio jets. 

\section{DISCUSSIONS AND SUMMARY}{\label{sum}}

To investigate spectral properties of YRGs and compare them with radio properties,
we construct a sample of 34 YRGs at relatively low redshift ($z < 0.4$)
for measuring narrow emission-line properties, \mbh, and Eddington ratio.
We determined \mbh\ from the width and luminosity of the broad \Ha\ line
using single-epoch mass estimators for Type 1 (broad-line) AGNs,
or from the measured stellar velocity
dispersion using the \msigma\ relation for Type 2 (narrow-line) AGNs.
The estimated \mbh\ ranges from 10$^{7.0}$ to 10$^{9.2}$ \msun, indicating YRGs have
relatively massive BHs, similar to the large-scale radio galaxies.

Based on the narrow emission-line flux ratios (e.g. \oiii/\Hb, [N {\sc ii}]/\Ha, [S {\sc ii}]/\Ha, and \oi/\Ha), 
we classified YRGs as HEG and LEG.
Most of Type 1 AGNs belong to HEGs, while Type 2 AGNs are composed of HEGs and LEGs.
We find that the Eddington ratio of HEGs is higher by $\sim$1.0 dex than that of LEGs,
using the \Ha\ line luminosity as a proxy for AGN bolometric luminosity. 
The difference in Eddington ratios and comparison with photoionization models
suggest that HEGs are high Eddington ratio AGNs with an optically thick accretion 
disk, which are similar to QSOs or Seyfert 1 galaxies, while LEGs have lower Eddington ratios 
with radiatively inefficient accretion flow. This interpretation is similar to
the division between Seyfert galaxies and LINERs in RQ AGNs 
(Kewley \etal~2006; Ho 2008), suggesting that YRGs have a various range of accretion activities
over 2-3 orders of magnitude in the Eddington ratio.

Kawakatu \etal~(2009) investigated whether the optical narrow emission-line 
ratios of YRGs are systematically different from those of RQ Seyfert 
2 galaxies by comparing the observed line ratios (e.g., [O {\sc i}]/[O {\sc iii}] and [O {\sc ii}]/[O {\sc iii}]) with photoionization models.
Using a limited sample of YRGs, they concluded that YRGs favor SED without a strong 
BBB, i.e., optically thin advection-dominated accretion flow, 
while RQ AGNs are consistent with the models adopting SED with a strong BBB, 
i.e., a geometrically thin, optically thick disk. 
In this study with an enlarged sample including Type 1 AGNs with higher Eddington ratios,
we find that there are various levels of accretion activity in YRGs and that both
SEDs with/without BBB are required to reproduce the observed line flux ratios
of YRGs. 

Low luminosity AGNs, i.e., LINERs, generally tend to be radio-loud (Ho 1999, 
Terashima \& Wilson 2003), implying that radio activity may be related with 
radiatively inefficient accretion flow, similar to the low-state in X-ray binaries 
(McClintock \& Remillard 2006).
In the case of YRGs, we find a large range of Eddington ratios, including
HEGs with high accretion power and Seyfert-like emission-line flux ratios.
Thus, the connection between radio jet and radiatively inefficient 
accretion flow is not strong in YRGs. Instead, YRGs are probably
composed of heterogeneous objects representing various accretion states.

By comparing narrow emission-line properties with radio luminosity and jet size,
we investigated the disk-jet connection in YRGs. 
The \oiii\ and \Ha\ line luminosities show broad correlations with the radio core luminosity,
indicating that accretion and radio activities in YRGs are fundamentally linked.
However, at fixed radio luminosity, HEGs have higher line luminosities (particularly for \oiii)
than LEGs, indicating that HEGs have higher accretion activity than LEGs for a given radio activity.
These results may suggest that at a given radio activity there is a continuous
distribution of accretion powers due to various mass accretion rate.

\acknowledgments

We thank the anonymous referee for constructive suggestions.
This work was supported by the Korea Astronomy and Space Science Institute (KASI) grant funded by the Korea government (MEST).
D.H.S acknowledges the support of the National Research Foundation of Korea Grant funded by the Korean Government (Ministry of Education, Science and Technology) [NRF-2010-355-C00026]. J.H.W acknowledges the support by the National Research Foundation of Korea (NRF) grant funded by the Korea government (MEST) (No. 2012-006087).
This work was supported in part by Ministry of Education, Culture, Sports, Science, and Technology (MEXT) Research Activity Start-up 2284007 (N.K.).

\end{document}